\documentclass[preprint]{aastex}
\begin{document}
\def\eg{{\it e.g.}}
\newcommand{\tnm}[1]{\tablenotemark{#1}}
\newbox\grsign
\setbox\grsign=\hbox{$>$}
\newdimen\grdimen
\grdimen=\ht\grsign
\newbox\simlessbox
\newbox\simgreatbox
\setbox\simgreatbox=\hbox{\raise.5ex\hbox{$>$}\llap
     {\lower.5ex\hbox{$\sim$}}}\ht1=\grdimen\dp1=0pt
\setbox\simlessbox=\hbox{\raise.5ex\hbox{$<$}\llap
     {\lower.5ex\hbox{$\sim$}}}\ht2=\grdimen\dp2=0pt
\def\simgreat{\mathrel{\copy\simgreatbox}}
\def\simless{\mathrel{\copy\simlessbox}}

\title{Young Clusters in the Nuclear Starburst of M~83\altaffilmark{1}}

\author{Jason Harris}
\affil{Space Telescope Science Institute}
\affil{3700 San Martin Dr., Baltimore, MD, 21218}
\affil{E-Mail: jharris@stsci.edu}
\author{Daniela Calzetti}
\affil{Space Telescope Science Institute}
\affil{3700 San Martin Dr., Baltimore, MD, 21218}
\affil{E-Mail: calzetti@stsci.edu}
\author{John S. Gallagher III}
\affil{Dept. of Astronomy, University of Wisconsin-Madison}
\affil{475 North Charter Street, Madison, WI 53706}
\affil{E-Mail: jsg@astro.wisc.edu}
\author{Christopher J. Conselice and Denise A. Smith\altaffilmark{2}}
\affil{Space Telescope Science Institute}
\affil{3700 San Martin Dr., Baltimore, MD, 21218}
\affil{consel@stsci.edu, dsmith@stsci.edu}

\altaffiltext{1}{Based on observations obtained with the NASA/ESA {\it
 Hubble
Space Telescope} at the Space Telescope Science Institute, which is
 operated
by the Association of Universities for Research in Astronomy, Inc.,
 under
NASA contract NAS5-26555.}

\altaffiltext{2}{Computer Sciences Corporation}

\begin{abstract}
We present a photometric catalog of 45 massive star clusters in the
nuclear starburst of M~83 (NGC~5236), observed with the Hubble Space
Telescope WFPC2, in both broad-band (F300W, F547M, and F814W) and
narrow-band (F656N and F487N) filters.  By comparing the photometry to
theoretical population synthesis models, we estimate the age and mass
of each cluster.  We find that over 75\% of the star clusters more
massive than $2\times10^4$~M$\odot$ in the central 300~pc of M~83 are
less than 10~Myr old.  Among the clusters younger than 10~Myr and more
massive than $5\times10^3$~M$\odot$, 70\% are between 5 and 7~Myr old.
We list an additional 330 clusters that are detected in our F300W
images, but not in the shallower F547M and F814W images.  The clusters
are distributed throughout a semicircular annulus that identifies
the active region in the galaxy core, between 50 and 130~pc from the
optical center of M~83.  Clusters younger than 5~Myr are
preferentially found along the perimeter of the semicircular annulus.
We suggest that the 5--7~Myr population has evacuated much of the
interstellar material from the active ringlet region, and that star
formation is continuing along the edges of the region.
\end{abstract}


\section{Introduction}\label{sec:intro}

A significant fraction of star formation in the universe may have
occurred in high-intensity bursts.  In the local universe, starburst
galaxies and spiral galaxies contribute roughly equally to the total
number of massive stars, despite the fact that spirals are much more
numerous \citep{hec98}.  At intermediate redshifts, the faint blue
galaxy population may be composed of dwarf galaxies undergoing
intense, rapidly-evolving starbursts \citep[\eg, ][]{col94, pd95}.
Finally, the Lyman-break galaxies at $z\sim3$ show evidence for
intense star formation \citep{ste96, ste99}.  Because starbursts are a
significant source of massive stars, from the local universe out to
$z\sim3$, they are a fundamental part of understanding the star
formation history of galaxies, and the chemical enrichment of the
universe.

Local starbursts make excellent laboratories for unraveling the
starburst mechanism, because their proximity affords superb spatial
resolution.  Local starbursts are quite similar to high-redshift
(z$\gtrsim$2.5) active galaxies in star formation rate \citep{meu97,
ken98}, UV colors \citep{meu97, mhc99, as00}, and spectral morphology
\citep{ste96, low97, pet00}, although the high-redshift galaxies tend
to have higher surface brightness than local starbursts \citep{wee98}.
Because of the many similarities, detailed photometric and
spectroscopic studies of local starbursts can constrain the processes
which govern star formation in active galaxies, and provide a
foundation for understanding the high-redshift objects.  In
particular, we wish to address the following questions: (1) Are
starburst properties determined by global properties of the host
galaxy? (2) What are the mechanisms that sustain starbursts?  (3) How
long do starbursts last?  (4) Does star formation propagate in
starburst galaxies? What is the propagation mechanism? (5) Do most
stars form in clusters (which then dissolve to form the field
population), or are there distinct modes of star formation for forming
clusters and field stars?

We are undertaking a study of local starburst galaxies to examine
these questions.  The subject of this paper is our photometric study
of M~83 (NGC~5236), with an emphasis on its star clusters.  M~83 is a
nearby \citep[3.7~Mpc,\ ][]{dev79} starburst galaxy.   It has a close
dynamical companion in NGC~5253 \citep{rog74}; these two very
different galaxies contain similarly intense nuclear starbursts
\citep{cal97b, cal99, tre01}.  The central starburst in M~83 spans
20~arcsec ($\sim360$ pc at 3.7~Mpc), and presents a relatively complex
morphology \citep[see\ ][]{gal91, elm98}. In contrast to the
centrally-concentrated starburst of NGC~5253, the optically-detected
starburst in M~83 is contained in a semi-circular annulus between 3''
and 7'' (54~pc and 126~pc at 3.7~Mpc distance) from M~83's optical
center \citep{tha00}.  This ``active ringlet'' region contains almost
400 star clusters brighter than $m_{F300W}=18$ mag, and 45 of these
are bright enough to be detected in the shallower F547M and F814W
images.

The nuclear region of M83 has several unusual properties, so we
are not surprised to find that it hosts a complex starburst region.
Located within the main bar of M83, the center of the galaxy appears to
contain a well-defined nuclear subsystem.  A bright optical nucleus
is offset from the center of the outer isophotes opposite to the
starburst ringlet \citep{gal91}, and a nuclear bar may separate these
two objects \citep{elm98}. Dense molecular gas is present and generally
concentrated to the north of the starburst ringlet \citep{ib01}, perhaps
a result of material collecting around an inner Lindblad resonance, as
suggested by \citeauthor{gal91}. Near infrared spectroscopy by
\cite{pdw97} confirms that young populations are offset from the
optically visible nucleus, corresponding to the optical starburst
ringlet. \cite{tha00} find that the optically visible nucleus, as well
as a second obscured nucleus at the center of symmetry, are dominated by
older red giant stars.

The picture emerging from these observations is of a complicated
situation, where the absence of axisymmetry is playing a role
in shaping ongoing events. The probable double nucleus suggests
that another galaxy merged with M83 in the past; its nuclear
starburst was not necessarily triggered only by the interaction
with NGC~5253. Since orbital time scales are short, $<$10$^7$~yr
within the starburst ringlet, the asymmetric distribution of
star forming activity evidently survived for several
orbits to explain the observed variations in cluster
ages (see Section~\ref{sec:agemass}). Furthermore, given the small 
distances involved and the near rigid body inner rotation curve found by
\citeauthor{tha00}, propagation of star formation even at the relatively
low speed of 10~km~s$^{-1}$ suffices to produce a starburst on the scale
of the ringlet in $\sim$10 ~Myr. Possibly, the starburst
is a transitory flare-up of circum-nuclear star
formation, in which molecular cloud complexes such as those
now present around the nucleus are rapidly disrupted,
rather than a long term event. The age distribution of star
clusters can shed light on this issue.

The paper is organized as follows: we present the observations and
data reduction in Section~\ref{sec:data}; photometry of the stellar
clusters and dust corrections are described in Sections~\ref{sec:phot}
and \ref{sec:dust}, respectively; age and mass determinations of the
clusters are made in Section~\ref{sec:agemass}, while the
properties of the diffuse population in the starburst site are
described in Section~\ref{sec:diffuse}.  We discuss the results of our
study in Section~\ref{sec:discuss}, and summarize the paper in
Section~\ref{sec:summary}.

\section{Observations and Data Reduction}\label{sec:data}

\begin{figure}
\plotone{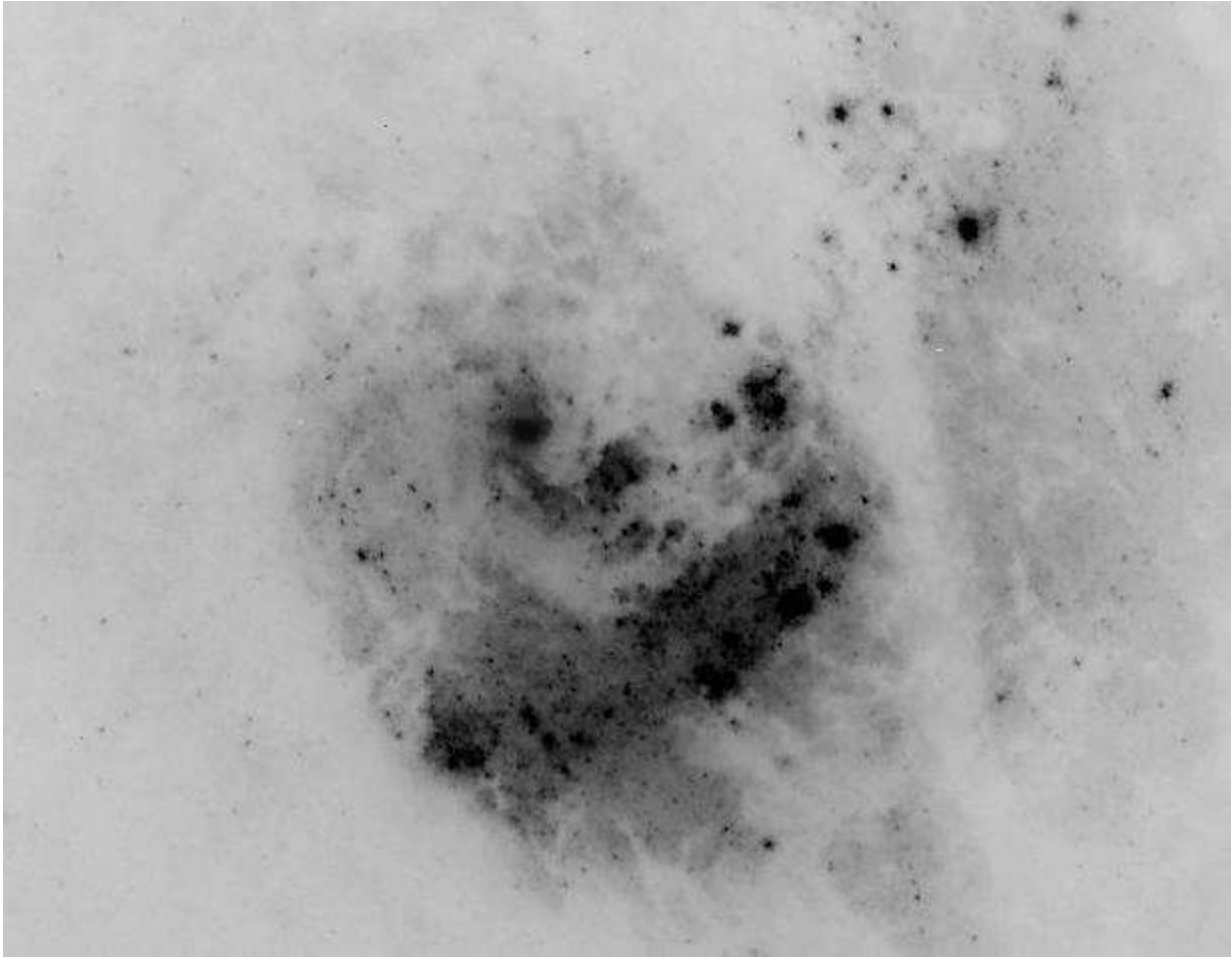}
\caption{F547M exposure of M~83.  North is at the top, and East
is at the left. (for a color composite image, see http://www.stsci.edu/~jharris/m83color.png) \label{fig:colorimage}}
\end{figure}

M~83 was observed with the Hubble Space Telescope (HST) Wide Field and
Planetary Camera 2 (WFPC2), in the broad-band filters F300W,
F547M and F814W and in the narrow-band filters F487N and F656N, in two
separate visits one~week apart during April-May, 2000 (see
Figure~\ref{fig:colorimage}). The log of the
exposures, together with the exposure times and the characteristics of
the filters are listed in Table~\ref{tab:exp}.  Images of the galaxy
in the F502N and F673N filters were also obtained during the same
visits and will be discussed in an accompanying paper
\citep{cal01}. The central starburst of the galaxy is about 20\arcsec~
in diameter and is perfectly matched with the PC1 chip
(36$^{\prime\prime}\times$36$^{\prime\prime}$), on which it was
centered. The telescope orientation was about $+$74.3 degrees off
North, and the rotation between the images obtained in the two visits
is small.

In each of the filters, two or three separate exposures were obtained
to aid in cosmic ray rejection. The F300W filter probes the galaxy
UV emission; the F814W filter is the WFPC2 equivalent I band; the
F547M filter is used here as a V-band equivalent, and was preferred to
the wider F555W filter because its bandpass excludes the strong
[OIII](5007~\AA) nebular emission from the galaxy.  The recession
velocity of M83 is sufficiently small (516~km\,s$^{-1}$) that the two
narrow band filters probe the hydrogen recombination emission lines
H$\beta$(4861~\AA) and H$\alpha$(6563~\AA), respectively.  The F656N
filter is narrow enough to exclude the redshifted [NII](6584~\AA)
emission line; however, the [NII](6548~\AA) emission line falls within
the filter's bandpass and represents about 15\% of the H$\alpha$ line
flux in the filter, as inferred from spectroscopic data
\citep{sck94}.

The data were reduced by the STScI calibration pipeline, which
includes flagging of bad pixels, A/D conversion, bias and dark current
subtraction, and flatfielding. Hot pixels were removed using the
STSDAS routine WARMPIX, which uses hot pixel information from dark
frames obtained around the time of the science observations to perform
the correction. The reduced images were registered to a common
position using linear shifts and small rotations; cosmic ray rejection
and co-addition were performed using the STSDAS routine CRREJ
\citep{wil96}, with a rejection threshold of 4~$\sigma$ for
the cosmic rays and 2.4~$\sigma$ for the adjacent pixels. The absolute
photometric calibration of the images is obtained from the zero-points
listed in HST Data Handbook \citep{hst97}, and have about 2\%--5\%
accuracy in the medium and broad-band filters \citep{cas97}. The
effect of contaminant buildup onto the WFPC2 window is negligible at
optical wavelengths, and was also negligible (less than 1\%) in the
F300W filter in this instance, because the observations were obtained
7 days after decontamination.  A correction of 2\% to the total counts
was applied to the narrow-band images to correct for the Charge
Transfer Efficiency (CTE) problem of WFPC2 \citep{ste98, whc99}.

Images of the nebular emission in H$\alpha+$[NII](6548~\AA) and in
H$\beta$ were constructed by subtracting the stellar continuum from
the narrow band images.  The image of the stellar continuum underlying
each of the lines was estimated from one of the broadband images
(F547M and F814W for F487N and F656N, respectively).  We used the
optical spectrum of \cite{skc95} as a reference for determining a
constant normalization factor applied to the broadband images.
Because of the galaxy's recession velocity, the H$\alpha$ line falls
close to the F656N red wing and a correction of
about 25\% was necessary to recover the full line flux; the line flux
was then corrected for the average value of the [NII](6548~\AA)
contribution \citep{sck94}, to obtain a final H$\alpha$ image. A 2\%
correction for the filter transmission curve was applied to the
H$\beta$ image. The effect of the underlying stellar absorption is
larger for the weak H$\beta$ emission than for H$\alpha$. Using red
stellar objects to scale the continuum image to the narrow-band image
can introduce a bias in the continuum subtraction of the blue stars,
depending on the red stellar population. For example, an F/G type
population has a relatively high H$\beta$ absorption equivalent width,
which would lead to undersubtraction of the OB stars continuum; on the
other hand, a K-type population with small Balmer equivalent widths
(EWs) would lead to over-subtraction of the blue stellar continuum. In
this case, a default correction of 2~\AA~ for the underlying stellar
absorption was applied to the H$\beta$ image, but individual cases
will be evaluated independently in the following analysis. The
continuum-subtraction and calibration of the emission line images were
cross-checked against the spectroscopic data of \cite{skc95}. Fluxes
were extracted from the images using the same aperture size and
orientation of the spectrum (10\arcsec$\times$20\arcsec, with the
largest size along the E-W direction). The aperture was centered at the
peak of the optical emission in the images (about the center of the PC-1
chip).  The flux densities as measured from the images are only a few
percent different than the values obtained from the spectrum convolved
with the WFPC2 filters bandpasses, confirming the accuracy of our
extraction.

\section{Star Cluster Photometry}\label{sec:phot}

Crowding in the starburst region of M~83 complicates the
identification of star clusters, and the accurate extraction of their
photometry.  We opted for profile-fitting photometry, rather than
aperture photometry, as the former allows us to disentangle partially
overlapping clusters and derive accurate total fluxes in each band.
In addition, profile fitting allows one to empirically check the
accuracy of the local background determination (a major source of
uncertainty in aperture photometry).  If the background of an object
is poorly determined, it will leave a circular pixel-value
discontinuity in the residual object-subtracted image.  We can then
iterate the fit until the residual image is smooth.

Profile fitting is typically applied to unresolved objects, which are
well-described by a single Point-Spread Function (PSF) across the
entire image or field.  The star clusters in M~83 are resolved in the
high angular resolution HST images; in the case of resolved objects,
PSF fitting can only be applied if the object profiles are uniform.
However, the clusters in our sample have profiles that vary in
significant and complex ways.  The half-light radius varies, and some
of the clusters have a sharp core feature superimposed on a broader
profile. As an added complication, the core feature can have a small,
random offset from the centroid of the underlying profile.  To
circumvent these problems, we convolve each of the images with a small
Gaussian kernel ($\sigma=2$ pixels), in order to get sufficiently
uniform cluster profiles that can be well-fit by a single PSF model.
The Gaussian kernel is large enough to homogenize the cluster
profiles, and small enough that blending of clusters is not a concern
(2 pixels corresponds to less than 2 pc at the distance of M~83).

\begin{figure}
\plotone{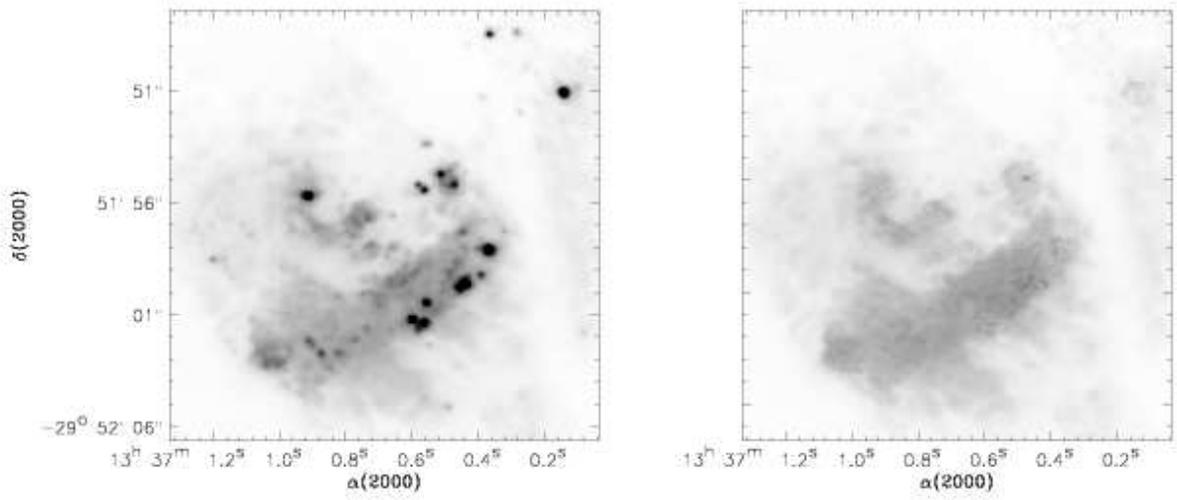}
\caption{{\bf Left:} the F547M image of the central starburst in M~83,
convolved with a $\sigma=2$ pixel Gaussian kernel.  North is at the top, and
east is to the left.  The bright peak near 13$^h$ 37$^m$ 0.9$^s$,
-$29^o$ 51' 55'' is the optical nucleus of the galaxy.  The
prominent dust lane can also be seen running north-south, in the
western part of the region. {\bf Right:} the same image, after all
detected objects have been fit and subtracted by DAOPHOT and ALLSTAR.
\label{fig:datimage}}
\end{figure}

We use DAOPHOT \citep{ste87} to perform iterative profile fitting on
the Gaussian-convolved images.  Objects are detected using DAOFIND,
with a $10\sigma$ detection threshold.  We obtain preliminary
photometry with PHOT, using a 3-pixel aperture.  Around 12 PSF objects
are selected with PSTSELECT, and the best-fit profile is determined
using PSF with a PSF radius of 11 pixels, and a fitting radius of 3
pixels.  We allow PSF to use all available functions in modeling
the profiles (Gaussian, Lorentz, Moffat, and Penny functions are
attempted).  The PSF is then used to subtract objects from the image
using SUBSTAR.  We examine the subtracted image for undetected objects
near the PSF objects (which are added to the list of objects to be
subtracted), and for poorly-fit PSF objects (which are removed from
the list of PSF objects).  We then run PSF again on an image in which
objects near the PSF objects have been subtracted.  The images require
between three and six such iterations to obtain cleanly subtracted
images (see Figure \ref{fig:datimage}).  When we have obtained a
satisfactory PSF model, we use ALLSTAR to obtain profile photometry of
all detected objects.  Since the magnitudes returned by the
ALLSTAR routine are normalized to a 3-pixel aperture, we
need to determine the aperture correction for each image to obtain
total magnitudes.

\begin{figure}
\plotone{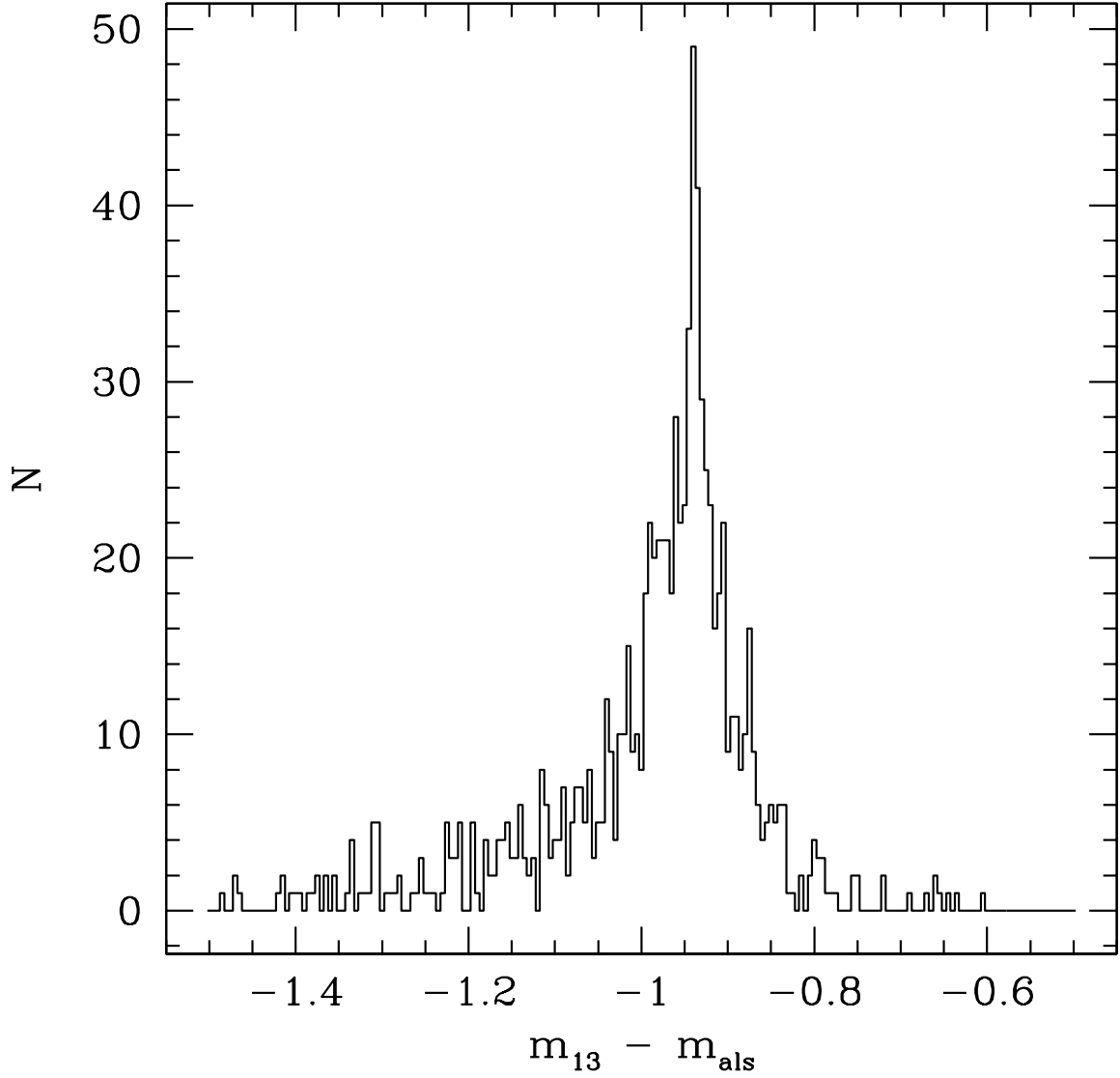}
\caption{A histogram of the aperture corrections derived from artificial
clusters added to isolated regions of the F547M image.  The aperture
correction is defined as the difference between the 13-pixel aperture
magnitude and the allstar magnitude.  We determine the mean
aperture correction by fitting a Gaussian function to the histogram peak,
and adopt the width of the peak as the contribution of the aperture
correction to the photometric errors. \label{fig:apcorr}}
\end{figure}

Aperture corrections are difficult to measure directly in
these images, because of the crowding of objects and the non-uniform
background light.  We therefore determine the aperture corrections
using artificial objects, added to the relatively empty image regions
outside the active ringlet region.  Artificial objects are
added to each image using the DAOPHOT ADDSTAR routine, which
constructs the objects from the image's best-fit PSF.  The artificial
objects are placed in a uniform grid, separated by 26 pixels in each
direction.  This ensures that the PSFs do not overlap, since the PSF
radius used is 11 pixels.  Once the artificial objects are added, we
select objects in isolated regions of the frame, and perform both
ALLSTAR photometry and concentric aperture photometry on the isolated
artificial objects.  The aperture correction is defined as the
difference between the 13-pixel radius aperture magnitude and the
ALLSTAR magnitude.  Figure~\ref{fig:apcorr} shows a histogram of the
aperture corrections from a series of F547M frames with artificial
objects added.  We fit a Gaussian to the peak to determine both the
mean aperture correction and its contribution to the photometric
errors.  We find that the scatter in the aperture correction dominates
the photometric error in all images (note that because the
distribution of aperture corrections has non-Gaussian tails, these
photometric errors may be somewhat underestimated).

We also use artificial-object tests to determine the completeness
limits in each image.  In this case, we add objects to the populated
regions of the image, so that the proper background levels and
crowding environments are sampled.  We estimate that the photometry is
90\% complete to 19.0 mag, 19.1 mag, and 19.6 mag, and 50\% complete
to 19.8 mag, 20.3 mag, and 20.4 mag for the F300W, F547M, and F814W
images, respectively.  Because of the finite number of artificial
objects used, the uncertainty in these completeness limits is
approximately 0.2 mag.  We convert these photometric completeness
limits to an age-dependent mass limit in Section~\ref{sec:discuss}.

These artificial object tests also allow us to test the accuracy of
profile-fitting photometry, relative to aperture photometry.  We find
that the profile-fitting photometry recovers the input artificial
magnitudes as accurately as does aperture photometry.  In fact,
profile-fitting is somewhat more accurate in highly crowded regions,
as we expected.

We produce a photometric catalog of star clusters by positionally
matching detected objects in the F300W, F547M and F814W images.  Only
objects which could be matched in all three images are included, and
the final catalog contains 45 star clusters (for a discussion of
potential selection effects, see section~7).  The astrometry and
broad/medium band photometry of the clusters in our sample are
presented in Table~\ref{tab:phot}.

Because the exposure time of the F300W image was twice as long as either
the F547M or F814W images (Table~\ref{tab:exp}), there are over 300
faint clusters detected in F300W, which were not detected in either
F547M or F814W.  64 of these clusters were detected in F547M, so we have
at least the $(m_{F300W} - m_{F547M})$ color for these objects.  We can
constrain the age of these clusters by examining the Starburst99 model
(see Figure~\ref{fig:2cd}): all model points with $(m_{F300W} -
m_{F547M})\simless -1.2$ are younger than 10 Myr.  We therefore
estimate that at least 42 of the 64 clusters detected in both F300W and
F547M are younger than 10 Myr.  We also attempted to constrain the ages
of the remaining F300W-detected clusters, by finding upper limits on
their $(m_{F300W} - m_{F547M})$ colors.  However, the color limits are
widely and uniformly distributed, so no significant constraint on the
ages was possible.  We present the astrometry and photometry of these
F300W-detected objects in Table~\ref{tab:uphot}.

\begin{figure}
\plotone{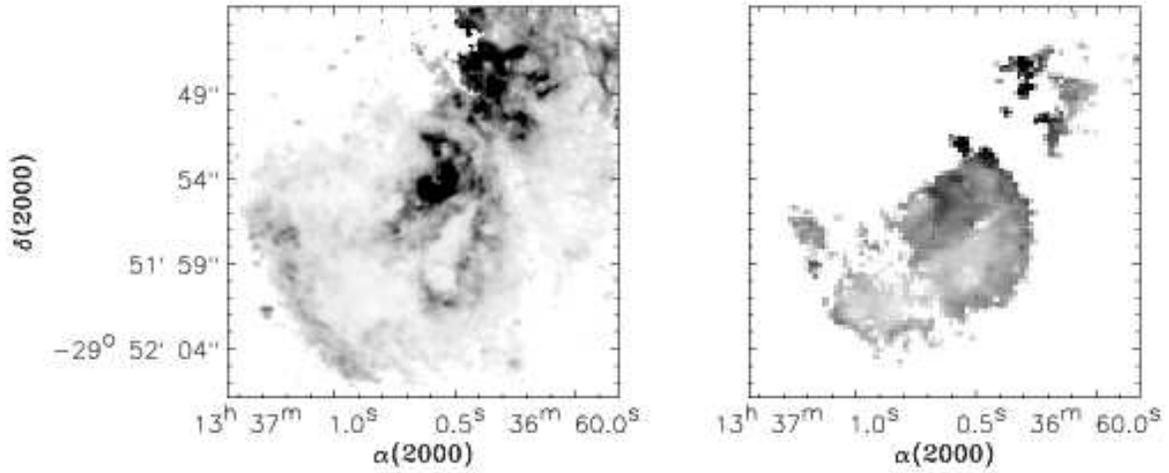}
\caption{{\bf Left:} A map of the EW(H$\alpha$) in the core of M~83.
Pixel values range from zero (white) to over 1000 (black). {\bf
Right:} A map of the $H\alpha/H\beta$ ratio for the observed region of
M~83.  Pixel values range from zero (white) to 10 (black).  We use
this ratio to estimate the dust extinction in each broadband image, as
explained in the text.  The large gap in the northwest region of the
map corresponds to the prominent dust lane, where signal in the
narrow-band images is lost due to heavy
extinction. \label{fig:halpha}}
\end{figure}

We perform a similar profile-fitting photometry procedure on the
narrow-band F656N (H$\alpha$) image (see Figure~\ref{fig:halpha}, left
panel).  Once the F656N magnitudes are determined, we calculate the
H$\alpha$ flux by using the same procedure described in
Section~\ref{sec:data} for the F656N image as a whole.  A
one-to-one match between optically-detected clusters and F656N
peaks cannot always be found in our images, because the ionized gas
associated with a cluster is not necessarily spatially coincident with
the cluster.  For example, evolving clusters can generate gas outflows
that clear the cluster's region of gas and dust, as observed in the
30~Doradus nebula by \cite{sco98} and \cite{wal99}. Therefore, we
choose to associate each peak detected in F656N with the nearest
 cluster, as
long as the cluster lies within a maximum search radius.  In this way, a
single cluster can accumulate the flux from several nearby F656N
peaks.  Because we are attempting to associate ionized gas with its
photoionizing source, we include all clusters detected in F300W in
this matching, regardless of whether they are in our final photometric
catalog.  In practice, it is difficult to choose an appropriate size
for the maximum search radius. We present two limiting cases: a smaller
radius of 5 pc, and a larger 12 pc radius.  These two strategies yield
identical F656N fluxes for all but three of the 45 bright clusters
in our sample.  There is one bright cluster (\#~28) for which no
matching F656N peak could be found.  We place an upper limit on
the F656N flux associated with this object by measuring the flux
in an 11-pixel aperture centered on the cluster in the peak-subtracted
F656N image.  The F656N flux is converted to an H$\alpha$ flux using
the procedure already described.  We derive the equivalent widths of
the H$\alpha$ emission (EW(H$\alpha$)) for each cluster from the
H$\alpha$ flux measurement, using the F814W flux associated with the
cluster to estimate the stellar continuum.  In
Section~\ref{sec:agemass}, we use the EW(H$\alpha$) values to help
constrain the cluster ages.  The EW(H$\alpha$) values are presented in
Table~\ref{tab:phot}.

\section{Dust Attenuation}\label{sec:dust}

\begin{figure}
\plotone{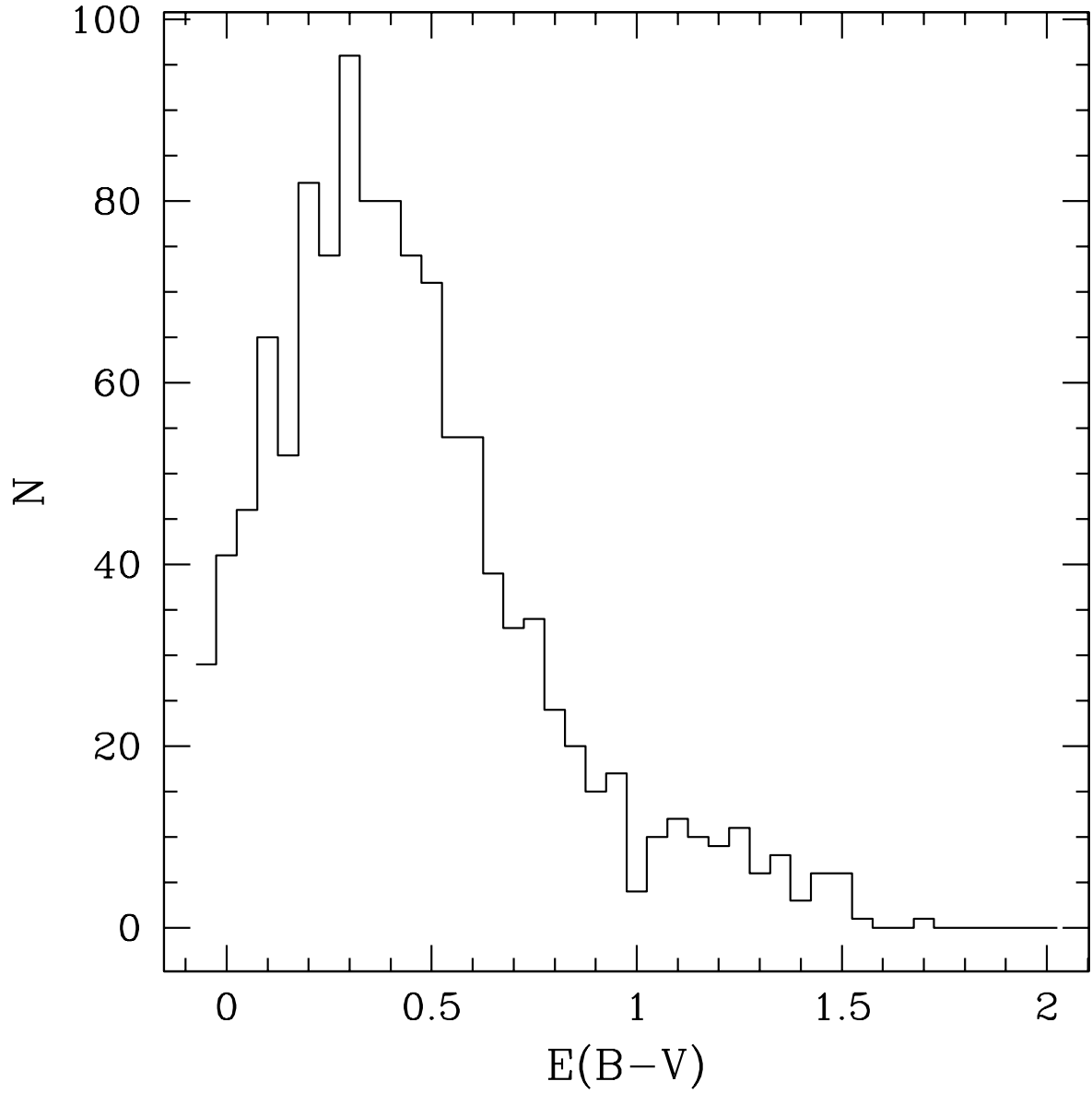}
\caption{The distribution of E(B-V) reddening values, as inferred from
the $H\alpha/H\beta$ ratio image (see Figure~\ref{fig:halpha}).
\label{fig:ebvdist}}
\end{figure}

We use the H$\alpha$ and H$\beta$ images to estimate dust extinction
corrections to the photometry.  The distribution of interstellar dust
is very complicated in the center of M~83, because of the presence of
an opaque dust lane \citep{tel93} and inhomogeneous dust and light
distributions.  We measure the ratio $R = F(H\alpha) / F(H\beta)$ at
each cluster location (see Figure~\ref{fig:halpha}, right panel), and
determine the $E(B-V)$ color excess using the standard formula for
foreground dust, appropriate for line emission in starburst galaxies
\citep{cal97}, and adopting an intrinsic ratio R$_{int}$=2.75
\citep{ost89}.  In Figure~\ref{fig:ebvdist}, we show the distribution
of inferred E(B-V) values for all pixels with $R>2.75$.  The
significant width of the distribution and its tail to high E(B-V)
values underscore the importance of determining an independent
reddening correction for each cluster.

Once the $E(B-V)$ color excess is determined for each cluster, we use
a two-component extinction model to deredden the photometry.  The
extinction model consists of a foreground Milky Way component
\citep[$E(B-V)_{MW}=0.06$,\ ][]{sfd98} and an internal component that
follows the starburst extinction curve \citep{cal94, cal00}:
\begin{eqnarray}
A_{300} = 3.07 \times (E(B-V)-0.06) + 5.59\times0.06\\
A_{547} = 1.79 \times (E(B-V)-0.06) + 3.10\times0.06\\
A_{814} = 1.14 \times (E(B-V)-0.06) + 1.79\times0.06
\end{eqnarray}

Four clusters lay outside the region for which reliable extinction
values can be determined using the $H\alpha~/~H\beta$ ratio, because
the signal-to-noise is too low in one or both of the narrow-band
images.  For these clusters, we correct the photometry along the
reddening vector in the color-color diagram (see Figure~\ref{fig:2cd})
so that the corrected photometry matches the colors predicted by
theoretical models (see next section).  If the cluster's reddening
vector intersects the model curve at more than one point, we use the
EW(H$\alpha$) associated with the cluster to constrain the cluster's
age, and use the age constraint to select among the possible
model points.

Correcting our images and photometry for the effects of dust using the
starburst obscuration curve \citep{cal94} provides the best agreement
overall between observations and models (see Section
\ref{sec:agemass}).  For instance, if we were to
use a model of foreground dust and a standard Milky Way extinction
curve \citep[\eg, ][]{sea79}, we would end up with reddening vectors
that are parallel to, but much longer than, those shown in Figure~4,
with the result that most reddening-corrected clusters would end up
with colors that are too blue to match any model.  More complex dust
geometries, such as a uniform mixture of dust and stars, could be
appropriate for some of our clusters (\eg, the youngest clusters, and
those close to the dust lane).  Mixed dust geometries produce less
steep reddening vectors ($\Delta(F547M-F814W)\sim
\Delta(F300W-F547M)\approx0.5$~mag) than foreground geometries,
resulting in age estimates that are $\approx$20\% older. In these
cases, age estimates from colors have been cross-checked against
similar estimates from the EW(H$\alpha$).

A general problem of using optical lines for deriving dust corrections
is the potential of underestimating the amount of dust present. For
foreground dust geometries this appears not to be the case, as larger
dust corrections than those adopted here would again produce clusters
colors that are too blue to be matched with the models. However, for
those few cases where a mixed dust/star geometry is not excluded,
there is a serious possibility of grossly underestimating the
cluster's intrinsic luminosity and, therefore, its mass. For these
cases, our mass estimates are treated as lower limits in
Table~\ref{tab:phot}.

\section{Determining Ages and Masses of the Star
Clusters}\label{sec:agemass}

We can estimate the ages and masses of the star clusters in our sample
by comparing our dust-corrected photometry and EW(H$\alpha$) to the
predicted photometry and EW(H$\alpha$) from population synthesis models.
For the comparison, we use the Starburst99 models \citep{lei99}, which
employ state-of-the-art isochrones and atmosphere calculations.  We
select an instantaneous burst population model covering the age range
1~Myr to 1~Gyr, with metallicity Z=0.040 \citep[to match that
of M~83,\ ][]{zar94, kob98} and a Salpeter IMF, covering stellar masses
between 1~$M\odot$ and 100~$M\odot$.  The models include flux from
continuum nebular emission for very young populations. We convolve the
model SEDs with the HST filter bandpass functions to obtain HST photometry
for each model point.  Finally, we perform a piecewise linear
interpolation between the original model points, to increase the
effective photometric and age resolutions of the model.

\begin{figure}
\plotone{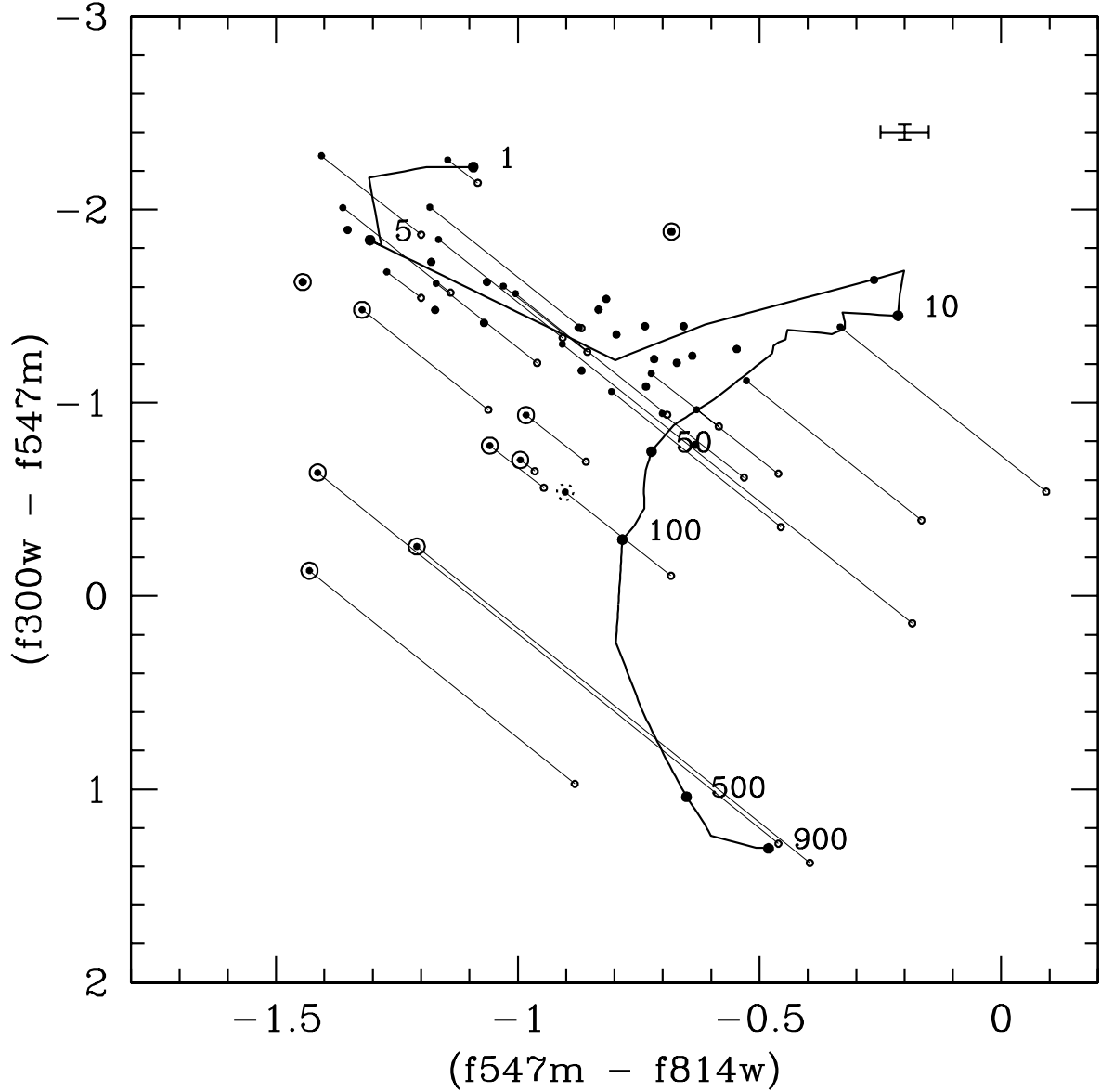}
\caption{A two-color diagram ($m_{547}-m_{814}$
vs. $m_{300}-m_{547}$), illustrating our cluster photometry.
Each cluster is represented by a pair of points, representing the
observed (open points) and extinction-corrected (solid points)
photometry.  The pair of points is connected by a line, for clarity.
The curve is the photometry of a Starburst99 population model with
Z=0.04, a Salpeter IMF, and an instantaneous burst star formation
history.  The model is a sequence in age; various ages are labeled (in
Myr) along the curve.  The mean photometric error for our cluster
sample is indicated by the error bars in the upper right
corner. The nine clusters whose extinction-corrected colors are more than
5$\sigma$ from any model point are indicated with large circles.
Object~\#4, whose photometry is {\it nearly} 5$\sigma$ from the model,
is indicated with a dotted circle.\label{fig:2cd}}
\end{figure}

In Figure~\ref{fig:2cd}, we present a two-color diagram, showing the
observed and dust-corrected photometry of our cluster sample.  The
Starburst99 model is also plotted; age increases along the model from
the upper left (bluer colors) to the lower right (redder colors).  For
the most part, our cluster sample follows the model prediction well,
although there are nine clusters whose photometry is strikingly
dissimilar from any point along the model.  We discuss these outliers
below.  We determine the ages of the clusters in our sample by simply
adopting the age of the model point whose colors are most similar to
the cluster's colors, and cross-checking the photometric age with the
cluster's EW(H$\alpha$).  We consider the age indeterminate if the
color-distance to the nearest model point exceeds $5\sigma$, where
$\sigma$ characterizes the photometric errors.

There are nine clusters in our sample for which no age
estimate could be made from the broadband colors.  All of these
clusters have EW(H$\alpha$) in excess of 80~\AA, indicating that they
are young (age $<$10~Myr).  With the exception of cluster \#~22, their
($m_{300} - m_{547}$) colors are 1-2 magnitudes redder than the locus
of young model points.  We also note that cluster \#~4, with a
photometric age of 72~Myr, has an EW(H$\alpha$)-derived age of
$\sim$6~Myr and occupies a similar position in the two-color diagram
as the nine outliers.  This object is particularly interesting,
because it corresponds to a bright IR/radio source \citep{gal91},
suggesting this object is even younger than our EW(H$\alpha$)
estimate.  We suspect the photometric anomalies may be
related to the dust correction, for at least some of these
clusters. Specifically, objects 5, 6, and 7 (the most anomalous
objects in Figure~\ref{fig:2cd}) each lie within the large dust lane
that bisects the observed region.  These objects are three of the four
most extincted objects in the sample, according to our
$H\alpha/H\beta$ measurements.  However, we note that no adopted
extinction value or dust model can make the photometry of these objects
consistent with the locus of young model points.  It therefore seems
likely that the dust geometry appropriate for these objects is much
more complex than the geometry implied by our adopted extinction curve
or other simple geometries.

There is a well-known uncertainty when applying population synthesis
models to relatively small populations like these clusters.  The
uncertainty is caused by the stochastic effect of luminous and rare
red supergiant stars.  The presence of these stars causes the "red loop"
visible in Figure \ref{fig:2cd}.  The shape and color of this loop
can change dramatically in a small population, because only a few red
supergiants are predicted, and random fluctuations in their actual
number have a large impact on the integrated colors.  However, because we
find few clusters in the 10--20 Myr age range (where red supergiants
dominate the integrated colors), we believe the impact of this uncertainty
on our results is minimal.

In addition, there is a possible bias in cluster ages derived from
EW(H$\alpha$) values, because the Starburst 99 models derive
EW(H$\alpha$) under the assumption that all far-UV photons are absorbed.
However, we believe this bias has only a small effect on our derived
ages.  Based on examining the diffused ionized gas in M~83,
\citet{cal99} find that only $\sim10\%$ of ionizing photons escape into
the general radiation field.  To test the impact of this photon leakage
on our cluster age estimates, we simulate full absorption of
ionizing photons by artificially increasing the EW(H$\alpha$) of the
clusters in our sample by 20\%.  On average, the H$\alpha$-derived
cluster ages decreased by only 8\% in this experiment.

\begin{figure}
\plotone{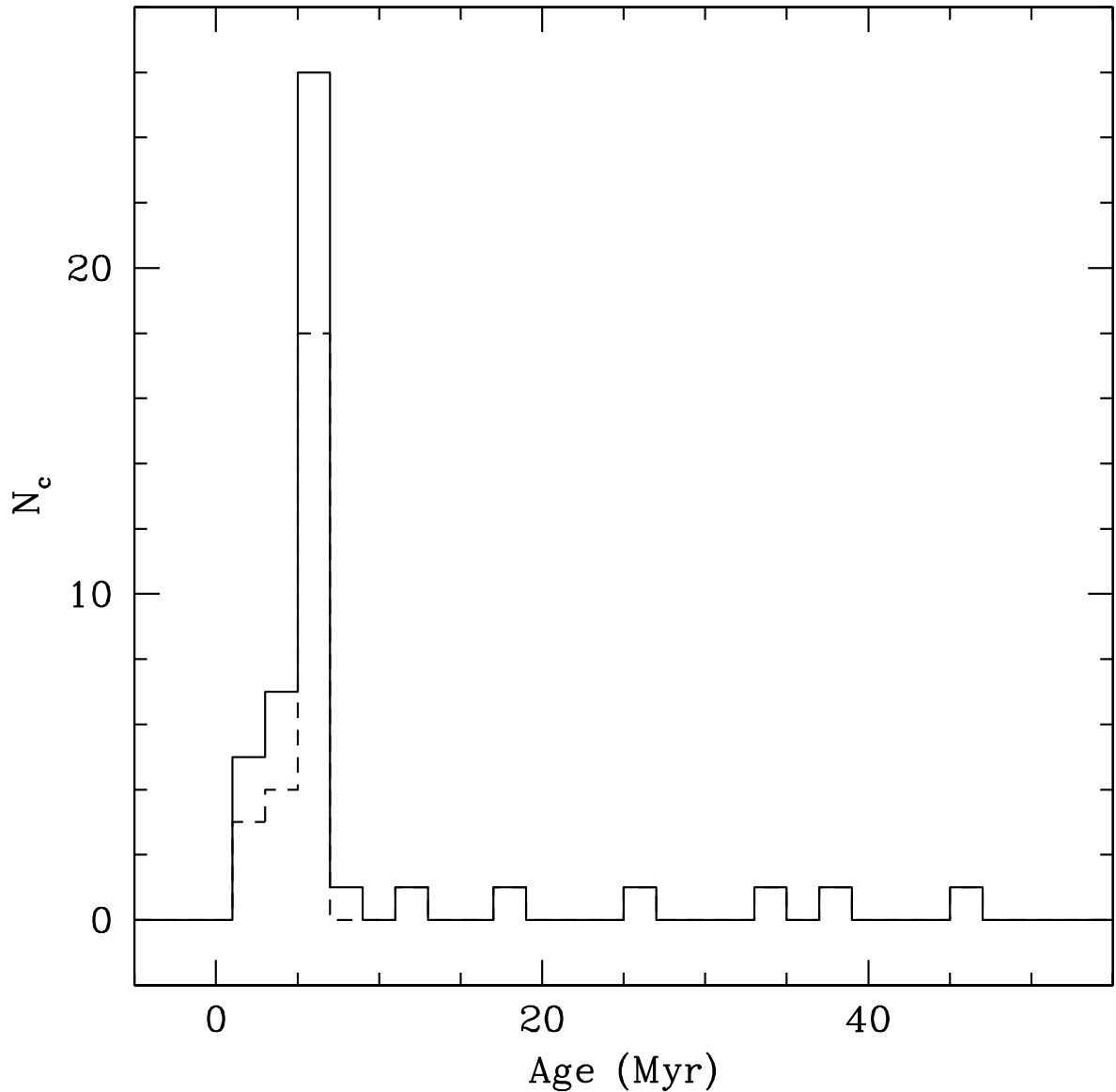}
\caption{The distribution of estimated ages for the clusters in our
sample. The dashed line is the distribution of ages, excluding
clusters with $M<2\times10^4$~M$\odot$.  Since our sample is complete
to this mass over the entire age range (see Figure~\ref{fig:massage}),
the fact that the strong peak at 5--7~Myr remains indicates that there
was an actual burst of activity at this time, as opposed to it being
the young end of a more continuous cluster formation episode.
\label{fig:agedist}}
\end{figure}

In Figure~\ref{fig:agedist}, we present the distribution of our
cluster age estimates from both photometry and EW(H$\alpha$).
Approximately 85\% of the 45 clusters in our bright sample are younger
than 10~Myr, and over half are in a narrow peak between 5 and 7~Myr.
Figure~\ref{fig:agedist} also shows the age distribution, excluding
clusters less massive than $2\times10^4$~M$\odot$.  The cluster sample
is complete above this mass over the entire age range of
Figure~\ref{fig:agedist} (see Figure~\ref{fig:massage}).  75\% of
these massive clusters are younger than 10~Myr.  The fact that the
peak between 5 and 7~Myr remains significant indicates that there was
an actual burst of cluster formation at that time; the peak is not
just the young end of a more continuous cluster formation episode,
whose older members have faded below our detection limit.  Nor is it
likely that such massive clusters would evaporate on timescales as
short as 10~Myr.

We determine the mass of each cluster in our sample by comparing the
observed, extinction-corrected flux in each filter with the predicted
flux of the corresponding model point.  The mass associated with the
model point is renormalized to that of a Salpeter mass function over the
range of stellar masses $0.35~M_\odot$ to $100~M_\odot$.  The lower mass
cutoff provides mass estimates consistent with a more realistic IMF
with a turnover at low masses \citep{kro01}.  The
flux ratio between a cluster and the corresponding model point is
assumed to be equal to their mass ratio.  For clusters with
photometrically-determined ages, we obtain three estimates of the mass
(one for each broadband filter), and adopt the mean value as the final
mass estimate. For ages inferred from EW(H$\alpha$), we derive the mass
from the F814W magnitude only, as this is the least affected by the
uncertain extinction correction. All but four of the clusters (\#~4, 7,
27, and 40 in Table~\ref{tab:phot}) have $M < 10^5 M_\odot$.  Thus, the
clusters in M~83 are less massive than those typically found in a major
merger galaxy like the Antennae \citep{fri99}.  We present two widely
divergent age and mass estimates for cluster \#~4, based on its
broadband photometry and its EW(H$\alpha$).  We consider the
photometric estimate less reliable, because the nearest model point in
the two-color diagram is 4$\sigma$ away, just barely within our
5$\sigma$ cut.  Also, its large EW(H$\alpha$) is convincing evidence
that the cluster must be young.  The masses of 10 of the clusters are
lower limits because we cannot discriminate a foreground dust geometry
from a mixed gas/dust geometry (see Section~\ref{sec:dust}).

\begin{figure}
\plotone{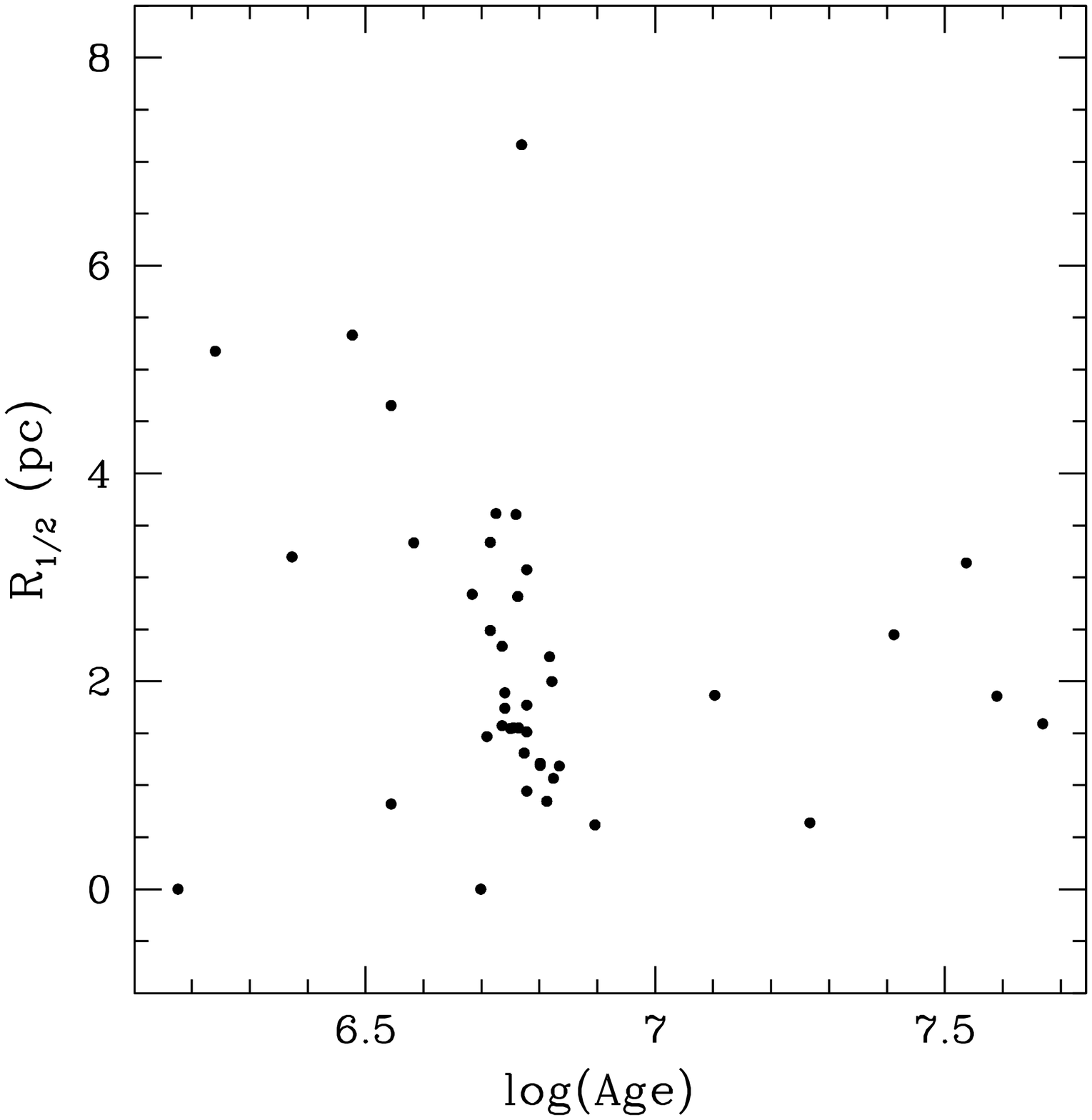}
\caption{The clusters' half-light radii plotted against their age.
The half-light radius is derived for each cluster from the F814W image.
\label{fig:radage}}
\end{figure}

We measure the half-light radii of the clusters in our sample using
concentric-aperture photometry of the unconvolved F814W image.  We
interpolate between the apertures to find the radius that encloses
half of the total flux (as measured by profile-fitting in the
convolved images).  Finally, we correct for the instrumental PSFs
half-light radius, adopting 1.6 pixels for the F814W PC chip PSF.
In Figure \ref{fig:radage}, we plot the clusters' ages against their
half-light radii.  There is some indication that the older clusters
tend to have smaller half-light radii, although the trend is very weak.

The results of these cluster measurements are summarized in
Table~\ref{tab:phot}, where the adopted color excess, the half-light
radius, the cluster mass estimate, and the age estimate are listed.

\section{The Diffuse Population}\label{sec:diffuse}

Having performed profile-fitting photometry on 45 clusters in our
images, we can use the ALLSTAR routine to produce images in which the
cluster light has been removed, leaving only the light of the diffuse
field population.  Note that for the purposes of this discussion, we
consider the 330 clusters detected only in F300W to be part of the
diffuse population.  We determine the mean ($m_{300} - m_{547}$) and
($m_{547}$ - $m_{814}$) colors of this diffuse stellar population by
performing a $5\times5$ blockaverage on the cluster-subtracted images,
and constructing relative magnitudes from the averaged pixel values.
The colors are simply the per-pixel differences in these relative
magnitudes.  We correct for extinction in each pixel using a similarly
blockaveraged $H\alpha/H\beta$ image, and applying the same
two-component extinction model as we use on the clusters. Finally, the
underlying galaxy's red population is removed from the diffuse
population colors by subtracting the average colors of the emission
surrounding the starburst.

\begin{figure}
\plotone{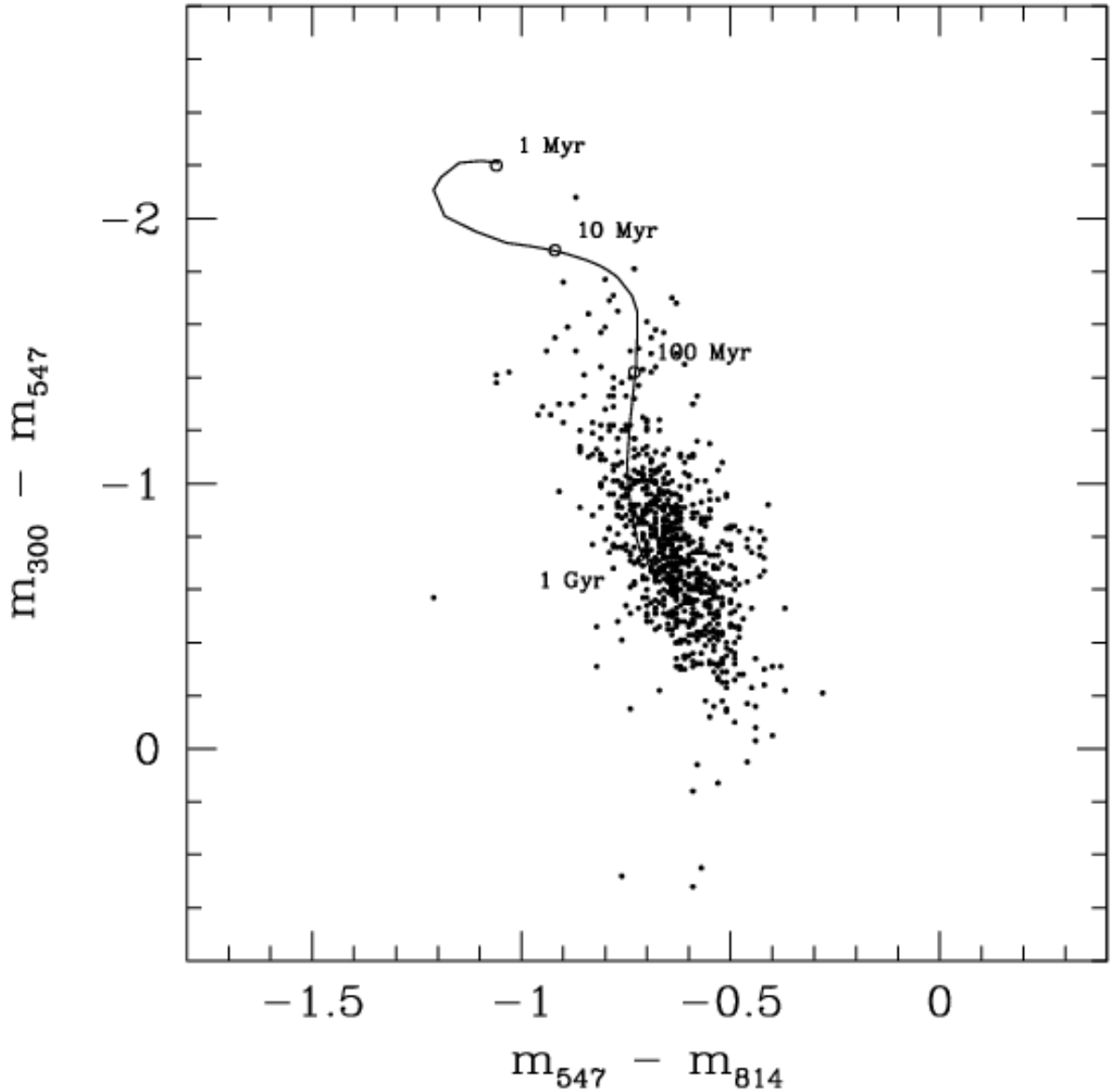}
\caption{The two-color diagram for the unresolved field population
in the M~83 starburst.  Each point represents the colors of a
$5\times5$ blockaveraged pixel in the cluster-subtracted images.
The curve is the photometry of a starburst99 population model with
Z=0.04, a Salpeter IMF, and a constant SFR. Various ages along the
model are labeled (in Myr), the ages correspond to the time at which
the constant star formation rate began. \label{fig:2cd-diffuse}}  
\end{figure}

\begin{figure}
\plotone{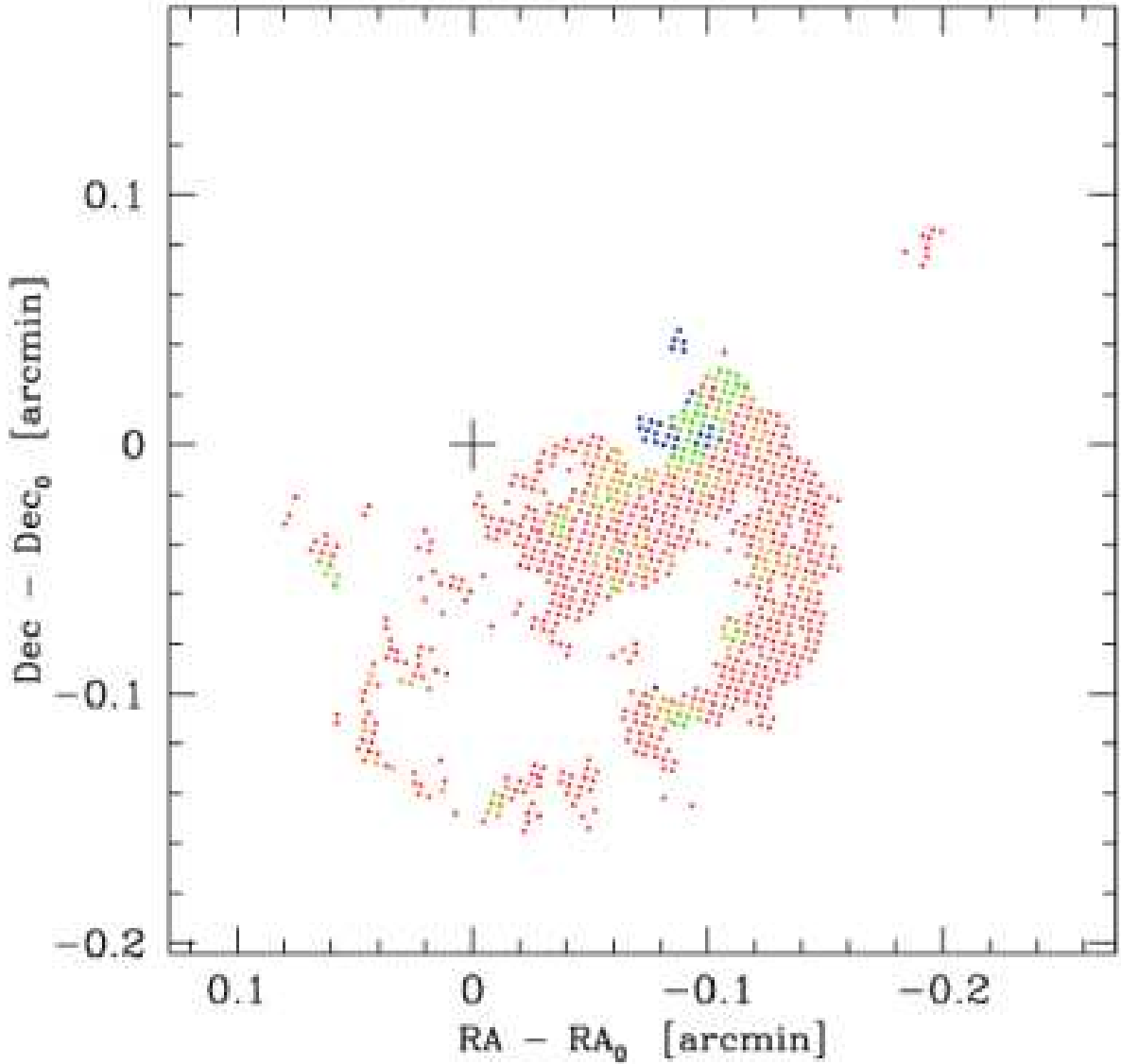}
\caption{A map of the ages of the oldest stars in the diffuse field
population, based on fitting the per-pixel colors of the field to a
constant star-formation rate model (see Figure \ref{fig:2cd-diffuse}).
The points are color-coded by age: 10--100~Myr (blue), 100--250~Myr
(green), 250--500~Myr (yellow), and 500--1000~Myr (red).  The field is
generally consistent with constant star formation since at least 1 Gyr
ago, although one region is better fit by a much younger initial age.
This region corresponds to a group of six young clusters, and one of
the two concentrations of large
EW(H$\alpha$). \label{fig:agemap-diffuse}}
\end{figure}

The two-color diagram for the diffuse stellar population is shown in
Figure~\ref{fig:2cd-diffuse}.  Only pixels with S/N $>10$ in each
filter are shown.  The curve represents a Starburst99 model for a
population with a constant  star-formation rate, a Salpeter IMF
covering stellar masses between 1 and 100~M$_\odot$, and metallicity
$Z=0.040$.  According to the model,
most of the field is consistent with a constant star-formation rate
starting 100~Myr to 1~Gyr ago, and all parts of the field contain
stars that are at least 10~Myr old.  This is similar to what was found
in NGC~5253 \citep{cal97, tre01}, although a much larger fraction of
the field population in M~83 contains stars older than 100~Myr.  In
Figure~\ref{fig:agemap-diffuse}, we show the map of the field
population's ``ages''.  It is important to note, these are {\bf not}
the average ages of the field populations.  The model assumes that the
star formation rate has been constant since some initial time $t_0$.
The map shows the distribution of $t_0$.  Pixels with large $t_0$
values could still contain very young populations; so large $t_0$
values are not necessarily inconsistent with the significant
EW(H$\alpha$) values throughout the region (Figure~3, left panel).
Without knowing how well the constant star formation rate model
fits the true star formation history, it is not possible to draw
strong conclusions about the distribution of ages of the diffuse
populations.

\begin{figure}
\plotone{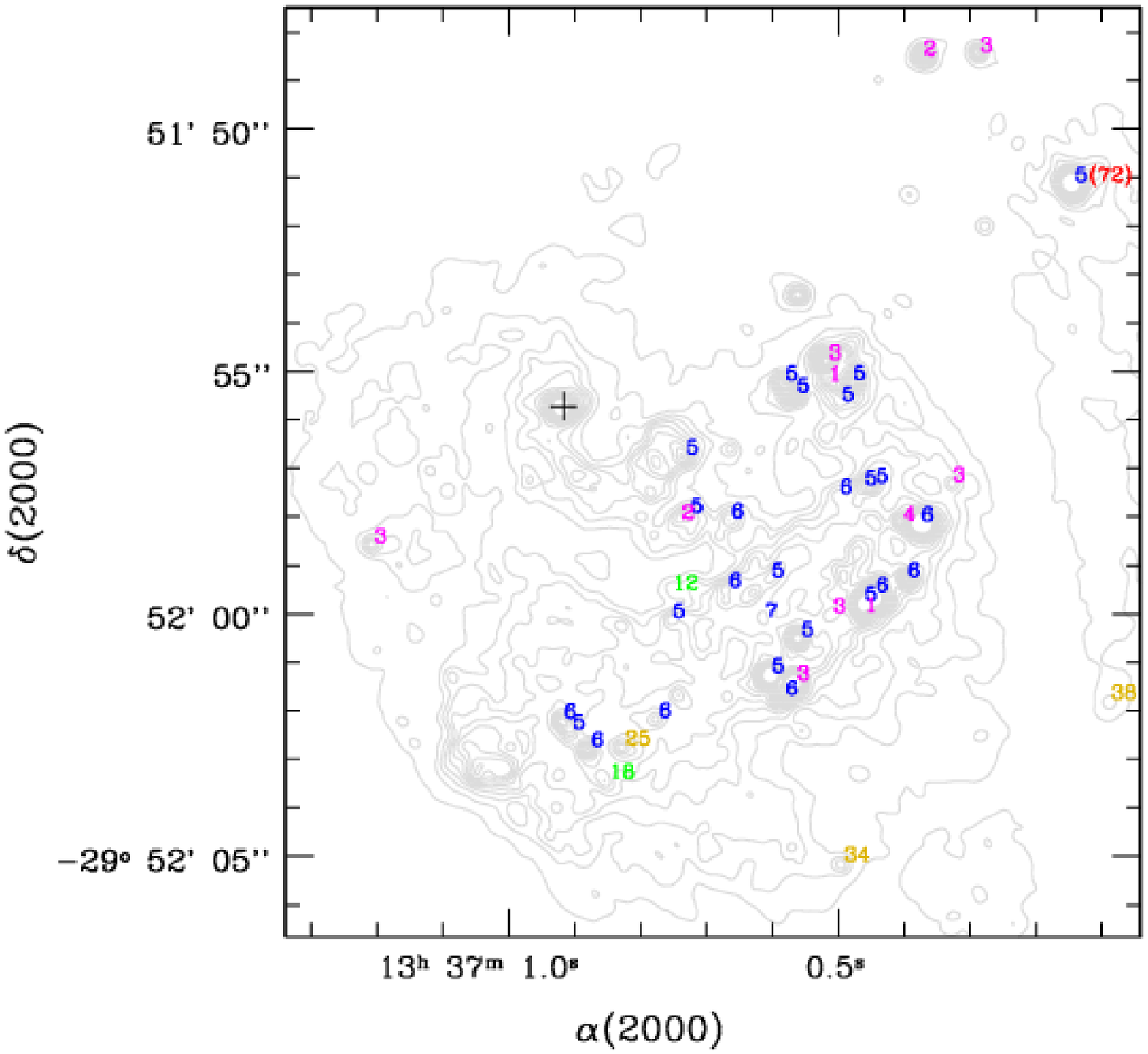}
\caption{A map of cluster positions in M~83.  Each cluster position
is labeled with the cluster's age to the nearest Myr.  In addition,
the ages are color-coded to guide the eye, according to the following
scheme: 0-5~Myr (magenta); 5-10~Myr (blue); 10-20~Myr (green);
20-40~Myr (yellow); and 40-100~Myr (red).  The grey contours trace the 
F547M image.  The center of M~83 is indicated with a cross, and the
coordinate axes are angular separation from this center. 
\label{fig:agemap}}
\end{figure}

With this caveat in mind, it is interesting to note the concentration
of low $t_0$ values to the west of M~83's optical center.  This area
also contains five young clusters (age$<$6~Myr; see
Figure~\ref{fig:agemap}), is characterized by large values of
EW(H$\alpha$) (Figure~3, left panel), and corresponds to the highest
concentration of H$\alpha$ flux. An area of 1.7$^{\prime\prime}$
radius around the peak nebular emission contains almost 20\% of all
the H$\alpha$ emission associated with the central starburst (the
total being $\sim$7.6~10$^{-12}$~erg~s$^{-1}$~cm$^{-2}$); the implied
star formation rate is 0.02~M$_{\odot}$~yr$^{-1}$ in an area of about
55$\times$55 pc$^2$, or about 10\%--15\% of the peak star formation
rate per unit area found in local starbursts \citep{meu97}.
This region represents the largest concentration of current star
formation in the core of M~83.  The second large concentration of
EW($H\alpha$) lies nearby, on the opposite side of the dust lane.
It is possible that the two observed H$\alpha$ peaks are in reality
one large star-forming area, bisected by the obscuring dust lane.
The area appears to coincide with the 10~$\mu$m emission peak reported
by \cite{tel93}. The entire area around the northern side of the dust
lane appears to be characterized by very recent, very active star
formation.

\section{Discussion: The Recent Cluster-Formation History of
 M~83}\label{sec:discuss}

The bulk of the visible recent cluster formation in the central region
of M~83 occurred southwest of the galaxy's optical center, in a
circumnuclear ringlet between 3'' and 7'' (54 pc and 126 pc) from the
nucleus (see Figure~\ref{fig:agemap}).  The majority of clusters more
massive than $2\times10^4$~M$\odot$ are between 5 and 7~Myr old, and
these clusters are found throughout the active ringlet
region.  In contrast, the extremely young ($<4$~Myr old) clusters are
found preferentially along the perimeter of the ringlet region.  Also,
the interior of the ringlet region contains mostly low-level H$\alpha$
emission, while its borders are lined by ridges of significant
(EW(H$\alpha)>100$) emission, culminating in the large H$\alpha$
concentration at the northern edge of the active ringlet.  The three
clusters with ages between 10~Myr and 30~Myr are all located around
the southernmost area of the ringlet, which also corresponds to the
larger of the two H$\alpha$ bubbles identifiable in the
ringlet. Clusters older than 30~Myr are all located outside the
ringlet;  however, we can conclude nothing about their origin from
their present location: assuming random velocities of order 50
km\,s$^{-1}$ \citep{cp89}, the crossing time of the observed region is
$\sim10$~Myr.

Star formation is clearly an ongoing process in the circumnuclear
region of M~83. The location of the youngest clusters and the
H$\alpha$ morphology suggest that star formation has propagated
from the interior of the ringlet to its perimeter. It appears that the
5--7~Myr population has evacuated interstellar material from most of
the active ringlet region; and that star formation is continuing along
the edges of the region, where the gas density is probably still
high. The age sequence from the presence of relatively old (up to
25~Myr) clusters in the southern edge of the ringlet to the peak of
recent star formation in the northern edge allow some tentative
inference that star formation has been ongoing in the southern part of
the ringlet for a longer time than in the northern part; possibly, it
has been propagating from one region to the other, a conclusion
already reached by \cite{gal91} from near-IR images and by
\citet{pdw97} from the CO and Br$\gamma$ emission of the region.
The starburst zone appears to be quite compact; propagation
at velocities of $\sim$10~km~s$^{-1}$ would spread star formation over
150~pc in only 15~Myr.

\begin{figure}
\plotone{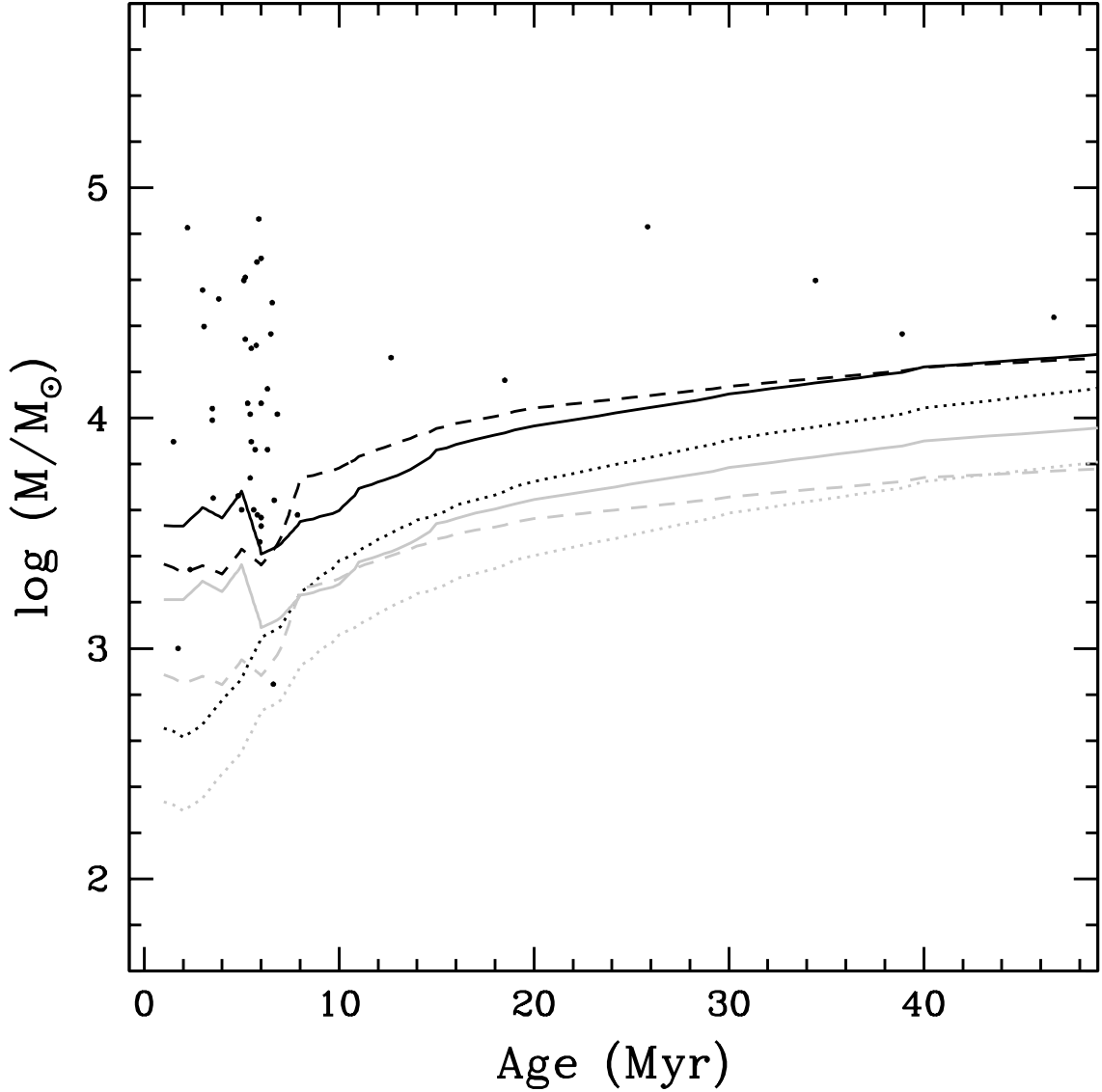}
\caption{The relationship between mass and age for clusters in our
sample.  The curves represent the minimum observable mass as a
function of age for the F300W (dotted lines), F547M (dashed
lines) and F814W (solid lines) images.  At each age, the minimum
observable mass is taken from the faintest Starburst99 model point
brighter than our 90\% (black lines) or 50\% (grey lines)
completeness limits in each image. \label{fig:massage}} 
\end{figure}

The relationship between masses and ages for the clusters in our
sample is shown in Figure~\ref{fig:massage}.  The 90\% completeness
limits impose a minimum observable mass as a function of age; this is
shown as a solid curve in Figure~\ref{fig:massage}.  Our sample
appears to be detection-limited at all ages.  Therefore, we cannot
say whether the lack of clusters older than 10 Myr with
M$<2\times10^4$~M$\odot$ is due to dynamical evaporation \citep[as
predicted for the centers of galaxies,\ ][]{kim00, por01}, or to simple
evolutionary fading of the lower-mass clusters.  However, clusters
more massive than $2\times10^4$~M$\odot$ are not expected to evaporate
on short timescales, so the fact that most massive clusters are
between 5 and 7 Myr old indicates that these clusters represent a true
burst of activity, and not the young end of a continuous cluster
formation rate.  This model of a sharp recent starburst is also
supported by the wholesale absence of dense gas across the ringlet, that
explains the abnormally blue FUV spectral energy distribution observed
in the center of M83 \citep{bon98}. Such a complete removal
of cool interstellar matter will not necessarily last , since
gas can refill the region during a sound crossing time, or about
30~Myr.

The conceptual models developed for the evolution of the center
of M83 by \cite{gal91} and \cite{tha00} appear to apply to these HST
data. Starbursts are associated with the rapid astration of molecular
clouds. The material in these clouds likely is fed into the nucleus by
the main bar, and possibly further shaped by an inner Lindblad resonance
and the influence of a nuclear bar.  When a sufficiently massive
cloud complex builds up, star formation occurs, and when sufficient
gas is removed, it becomes optically visible as is currently seen
in M83. The masses of the molecular clouds now present in M83
of $\sim 2 \times 10^7$ M$\odot$ \citep{ib01} could produce
a starburst comparable to that which we see, with a star formation
efficiency of $<$10\%. Since the cloud sizes are also a few hundred
pc, the duration of the event should be 10-30~Myr. Given the lack
of star clusters with ages of $>$50~Myr in our data, the typical
time span between events should be longer than this, which
still allows multiple bursts to occur in the last Gyr and
thereby fit the diffuse stellar population colors. If each event
adds about $\sim 2 \times 10^6$ M$\odot$ of new stars, then 10
events per Gyr would just about suffice to produce the mass
in the inner 100~pc of this galaxy. Starbursts may have become
a normal evolutionary event in M83.

\begin{figure}
\plotone{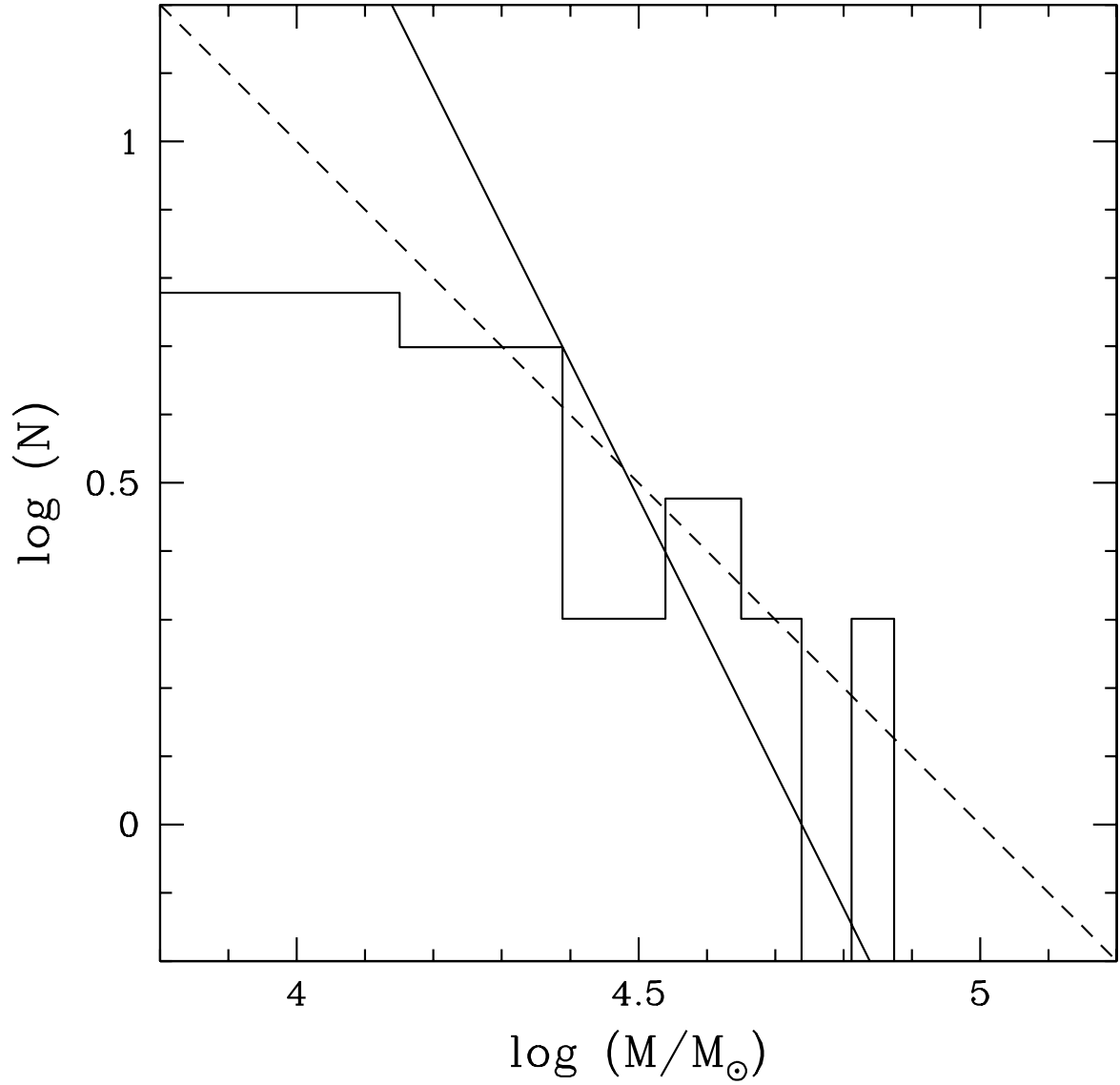}
\caption{The distribution of masses for young (age$<10$~Myr)
clusters in our sample.  The solid line represents a power law with 
$\alpha=-2.0$, which is similar to the mass function observed in
major-merger starbursts. The dashed line represents a power law with 
$\alpha=-1.0$.  \label{fig:imf}} 
\end{figure}

The distribution of masses of relatively young (age $<10$~Myr),
massive (mass $>10^4$~M$\odot$) clusters (Figure~\ref{fig:imf})
provides a look at the cluster mass function (MF) for the M~83
starburst.  The observed MF is reasonably consistent with the
power-law cluster MF observed in the major-merger starbursts NGC~1741
\citep{joh99} and NGC~4038/9 \citep{zf99} ($\Psi(M)\propto M^{-2.0}$).
However, the mass function is based on only 25 young, massive clusters
in our sample, and the dynamic range of masses is small, so the
slope is only weakly constrained.

\section{Summary}\label{sec:summary}

Our analysis of the HST WFPC2 images of M~83 indicates that this
galaxy experienced a very recent starburst with a peak of massive
cluster formation within the last 5--7~Myr, and that the starburst may
have triggered even more recent star formation activity along the
perimeter of the starburst region.

We measured both broadband (F300W, F547M, and F814W) and narrow-band
(F656N) photometry for 45 clusters in the central 300 pc of M~83. We
were able to determine the ages of 36 of these clusters by matching
their positions in a two-color diagram to a Starburst99 model, and
constrain the ages of the remaining nine clusters using the
EW(H$\alpha$). Of these clusters, 39 are younger than 10~Myr, and
25 are in the narrow age range 5--7~Myr.  Clusters in this age range
are distributed in a semicircular annulus (the active ringlet),
between 50 and 130 pc from the center of M~83.  We find that the five
clusters with ages younger than 4~Myr are found exclusively along the
edges of the active ringlet region, suggesting an outward propagation
of star formation. In addition, the presence of clusters older than
10~Myr in the southern end of the ringlet and of current, intense
star formation at the northern end suggests that star formation may
have started earlier in the southern region and propagated northward, 
in agreement with a similar inference by \citet{gal91} and
\citet{pdw97}.

The look-back time on the star formation history of the starburst in
M~83 is limited by the scant information provided by the few clusters
older than $\sim$10~Myr and by the diffuse population. The latter,
made mostly of a $>10$~Myr population, may be composed of the remnants
of evaporated star clusters.  The limited information available
precludes us from drawing strong conclusions about the origin of the
starburst in M~83.  The nearly identical time evolution of the
starbursts in M~83 and NGC~5253 suggests a causal connection for the
starbursts, such as a galactic interaction.  However, perigalacticon
in this system probably occurred between 1 and 2 Gyr ago \citep{rog74,
vdb80, cp89}; much older than the observed burst populations.  The
diffuse population in M~83 perhaps offers a glimpse of what remains of
star formation triggered directly by this encounter, since much of the
diffuse population is consistent with a constant star-formation rate
starting at least 1 Gyr ago.  It is perhaps more likely that the
present nuclear star forming activity in M~83 is driven and sustained
by gas inflow from its bar instability.  

\vskip 1in
\noindent Acknowledgements:
This work has been supported by the NASA LTSA grant NAG5-9173 and by the NASA HST grant GO-08234.01-A.

%Bibliography
%\bibliographystyle{apj}
%\bibliography{/home/jharris/bibtex/starburst}

\clearpage

%%Begin Tables
%\include{harrisj.tab1}
\begin{deluxetable}{cccccc}
\tablewidth{0pt}
\tablecaption{Log of the Exposures\tnm{a}\label{tab:exp}}
\tablehead{
\colhead{Exposure}           & \colhead{Obs. Date} & \colhead{Filter}      &
\colhead{Mean Wavelength}          & \colhead{Filter Bandpass}  &
\colhead{Exposure Time} \\
\colhead{Name}    & \colhead{DD-MM-YY} & \colhead{ }  & \colhead{(\AA)}  &
\colhead{(\AA)} & \colhead{(sec)} }
\startdata
U5BG0101 & 25-04-00 & F300W & 2943. & 735.8 & 700. \\
U5BG0102 & 25-04-00 & F300W & 2943. & 735.8 & 700. \\
U5BG0103 & 25-04-00 & F300W & 2943. & 735.8 & 700. \\
U5BG0104 & 25-04-00 & F547M & 5476. & 483.1 & 180. \\
U5BG0105 & 25-04-00 & F547M & 5476. & 483.1 & 350. \\
U5BG0106 & 25-04-00 & F547M & 5476. & 483.1 & 400. \\
U5BG0107 & 25-04-00 & F814W & 7921. & 1488.8 & 160. \\
U5BG0108 & 25-04-00 & F814W & 7921. & 1488.8 & 200. \\
U5BG0109 & 25-04-00 & F814W & 7921. & 1488.8 & 350. \\
U5BG0201 & 02-05-00 & F487N & 4865. & 25.8 & 1100. \\
U5BG0202 & 02-05-00 & F487N & 4865. & 25.8 & 1200. \\
U5BG0207 & 02-05-00 & F487N & 4865. & 25.8 & 1000. \\
U5BG0208 & 02-05-00 & F656N & 6564. & 21.4  & 600. \\
U5BG0209 & 02-05-00 & F656N & 6564. & 21.4  & 600. \\
\tablenotetext{a}{The nominal telescope pointing was 
$\alpha$(J2000)=13:37:00.75, $\delta$(J2000)=$-$29:51:57.0 in the
aperture PC1-FIX.}
\enddata
\end{deluxetable}

\clearpage

\begin{deluxetable}{rccccccccc}
\tabletypesize{\scriptsize}
\tablecolumns{10}
\tablewidth{0pt}
\tablecaption{Astrometry and Photometry of clusters in NGC 5236
\label{tab:phot}}
\tablehead{
   \colhead{ID} & \colhead{$\alpha$(2000)} & \colhead{$\delta$(2000)} & \colhead{$m_{F300W}$} & \colhead{$m_{F547M}$} & 
   \colhead{$m_{F814W}$} & \colhead{EW(H$\alpha$)\tnm{a}} & \colhead{E(B-V)} & \colhead{Mass} & \colhead{Age} \\

   \colhead{ } & \colhead{ } & \colhead{ } & \colhead{(STMAG)} & \colhead{(STMAG)} & \colhead{(STMAG)} & 
   \colhead{(\AA)} & \colhead{(mag)} & \colhead{($10^3 M\odot$)} & \colhead{(Myr)} \\
}

\startdata
 1 & 13:36:59.80 & -29:51:54.81 & 17.73 & 18.81 & 19.55       &  \nodata    & 0.00        &   49 &   6.00 \\
 2 & 13:36:59.96 & -29:51:46.34 & 18.42 & 19.20 & 19.84       &  \nodata    & 0.00        &   27 &  46.67 \\
 3 & 13:37:00.09 & -29:52:01.76 & 18.31 & 19.25 & 19.95       &  \nodata    & 0.20\tnm{b} &   23 &  38.89 \\
 4 & 13:37:00.14 & -29:51:51.08 & 14.86 & 15.40 & 16.30       &  125        & 0.28        &   48 &   5.79 \\
\nodata& \nodata & \nodata      &\nodata&\nodata&\nodata      & \nodata     & \nodata     &(1074)\tnm{c}&(72.5)\tnm{c} \\
 5 & 13:37:00.28 & -29:51:48.39 & 15.88 & 16.13 & 17.34       &  734/968    & 1.20        &   25\tnm{d} &   3.07\tnm{e} \\
 6 & 13:37:00.32 & -29:51:57.27 & 17.59 & 17.72 & 19.15       &  480        & 0.79        &  4.5\tnm{d} &   3.54\tnm{e} \\
 7 & 13:37:00.37 & -29:51:48.47 & 14.07 & 14.71 & 16.13       & 1052/1670   & 1.42        &   67\tnm{d} &   2.22\tnm{e} \\
 8 & 13:37:00.37 & -29:51:58.08 & 14.74 & 15.68 & 16.66       &   96        & 0.13        &   32 &   6.57\tnm{e} \\
 9 & 13:37:00.39 & -29:51:59.23 & 16.60 & 17.83 & 18.54       &   39        & 0.00        &   13 &   6.33 \\
10 & 13:37:00.40 & -29:51:58.06 & 17.48 & 18.26 & 19.32       &  325        & 0.11        &  4.6 &   4.83\tnm{e} \\
11 & 13:37:00.43 & -29:51:59.31 & 17.82 & 18.52 & 19.52       &  255        & 0.02        &  4.0 &   5.00\tnm{e} \\
12 & 13:37:00.44 & -29:51:57.29 & 17.00 & 18.54 & 19.35       & \nodata     & 0.00        &  7.3 &   5.69 \\
13 & 13:37:00.44 & -29:51:59.53 & 16.04 & 17.28 & 17.92       &   47        & 0.00        &   23 &   6.50 \\
14 & 13:37:00.46 & -29:51:57.34 & 17.66 & 19.08 & 20.15       & \nodata     & 0.00        &  4.0 &   5.62 \\
15 & 13:37:00.46 & -29:51:59.72 & 15.85 & 17.33 & 18.50       &  600        & 0.00        &   20 &   5.50 \\
16 & 13:37:00.46 & -29:51:59.94 & 16.03 & 18.29 & 19.43       & \nodata     & 0.05        &  7.9 &   1.50 \\
17 & 13:37:00.47 & -29:51:55.18 & 15.64 & 17.32\tnm{f}& 18.59 &  408        & 0.05        &   22 &   5.19 \\
18 & 13:37:00.49 & -29:51:55.61 & 16.86 & 18.42 & 19.43       & \nodata     & 0.43        &  7.9\tnm{d} &   5.50 \\
19 & 13:37:00.49 & -29:52:05.10 & 17.63 & 18.59 & 19.22       & 1279        & 0.20\tnm{b} &   40 &  34.44 \\
20 & 13:37:00.50 & -29:51:57.51 & 18.11 & 19.26 & 19.98       & \nodata     & 0.16        &  3.4 &   6.00 \\
21 & 13:37:00.51 & -29:51:55.19 & 15.97 & 17.98 & 19.17       &  228        & 0.43        &   11\tnm{d} &   3.50 \\
22 & 13:37:00.51 & -29:51:59.97 & 18.01 & 19.90 & 20.58       & 1907        & 0.00        &  1.0 &   1.74\tnm{e} \\
23 & 13:37:00.52 & -29:51:54.75 & 14.41 & 16.68 & 18.09       &  351        & 0.26        &   36 &   3.00 \\
24 & 13:37:00.56 & -29:51:55.43 & 14.93 & 16.77 & 17.94       &   17        & 0.34        &   40 &   5.12 \\
25 & 13:37:00.56 & -29:52:00.45 & 15.91 & 17.39 & 18.22       & 1365        & 0.00        &   21 &   5.75 \\
26 & 13:37:00.56 & -29:52:01.35 & 14.77 & 16.67 & 18.02       &   43        & 0.00        &   33 &   3.83 \\
27 & 13:37:00.58 & -29:51:55.18 & 14.62 & 15.92 & 16.83       &  264/807    & 1.00\tnm{b} &   73\tnm{d} &   5.88 \\
28 & 13:37:00.58 & -29:52:01.67 & 16.44 & 18.06 & 19.51       &   81\tnm{g} & 0.00        &  0.7 &   6.63\tnm{e} \\
29 & 13:37:00.60 & -29:51:59.23 & 18.13 & 19.48 & 20.28       & \nodata     & 0.00        &  2.9 &   5.94 \\
30 & 13:37:00.60 & -29:52:01.20 & 14.96 & 16.69 & 17.87       &   87        & 0.00        &   41 &   5.19 \\
31 & 13:37:00.61 & -29:52:00.05 & 18.30 & 19.93 & 20.20       & \nodata     & 0.00        &  3.8 &   7.87 \\
32 & 13:37:00.66 & -29:51:58.02 & 16.82 & 17.88 & 18.68       & \nodata     & 0.48        &   12\tnm{d} &   6.00 \\
33 & 13:37:00.67 & -29:51:59.44 & 17.94 & 19.11 & 19.98       &  218        & 0.00        &  3.7 &   6.00 \\
34 & 13:37:00.72 & -29:51:57.90 & 16.55 & 18.15 & 19.18       &  239        & 0.21        &   10 &   5.44 \\
35 & 13:37:00.73 & -29:51:56.70 & 16.40 & 18.01 & 19.18       & \nodata     & 0.26        &   12 &   5.31 \\
36 & 13:37:00.74 & -29:51:58.05 & 17.08 & 18.56 & 19.88       & 1467        & 0.34        &  2.2 &   2.36\tnm{e} \\
37 & 13:37:00.75 & -29:51:59.50 & 17.32 & 18.71 & 19.04       &  156        & 0.60\tnm{b} &   18\tnm{d} &  12.67 \\
38 & 13:37:00.75 & -29:52:00.07 & 17.80 & 19.19 & 20.06       &  118        & 0.00        &  3.8 &   5.81 \\
39 & 13:37:00.77 & -29:52:02.12 & 17.74 & 19.14 & 19.88       & \nodata     & 0.00        &  4.4 &   6.67 \\
40 & 13:37:00.82 & -29:52:02.70 & 16.72 & 17.83 & 18.36       &   13        & 0.50        &   68\tnm{d} &  25.83 \\
41 & 13:37:00.85 & -29:52:03.39 & 18.00 & 19.28 & 19.82       & \nodata     & 0.00        &   15 &  18.50 \\
42 & 13:37:00.88 & -29:52:02.73 & 16.86\tnm{f}& 18.25 & 18.91 &   16        & 0.00        &   10 &   6.83 \\
43 & 13:37:00.90 & -29:52:02.37 & 17.22 & 18.84 & 19.91       & \nodata     & 0.00        &  5.5 &   5.44 \\
44 & 13:37:00.91 & -29:52:02.14 & 17.28 & 18.49 & 19.16       &   46        & 0.00        &  7.3 &   6.33 \\
45 & 13:37:01.20 & -29:51:58.53 & 16.02 & 18.02 & 19.39       &  255        & 0.28        &  9.8 &   3.50 \\
\enddata

\tablenotetext{a}{EW(H$\alpha$) values estimated by flux-adding all H$\alpha$ peaks
                  within 5 and 12 pixels of the cluster position.  In the three cases
                  where the two radii yield different EW(H$\alpha$) values, both are presented.} 
\tablenotetext{b}{No reliable extinction estimate could be made from the 
                  $H\alpha/H\beta$ map.  We assume a reddening value that makes the
                  intrinsic cluster photometry consistent with a point on the
                  Starburst99 model.} 
\tablenotetext{c}{Probable anomalous cluster. Primary age and mass estimates are from EW(H$\alpha$); estimates 
                  from photometry are given in parentheses.} 
\tablenotetext{d}{We cannot exclude a mixed gas/dust geometry for objects with E(B-V)$\simgreat$0.37 mag.  
                  The mass presented is a lower limit, because it assumes the intrinsic luminosity is not 
                  underestimated.}
\tablenotetext{e}{Cluster has anomalous photometry.  Age and mass estimates are from EW(H$\alpha$).}
\tablenotetext{f}{The cluster was split into two detections in this filter.  We flux-added the two detections 
                  to get the total cluster flux.}
\tablenotetext{g}{Upper limit to EW(H$\alpha$) determined from cluster-centered aperture of peak-subtracted 
                  F656N image.  The age and mass estimates for this cluster are lower limits.}
\end{deluxetable}

\clearpage

\begin{deluxetable}{rcccccc}
\tabletypesize{\scriptsize}
\tablecolumns{7}
\tablewidth{320pt}
\tablecaption{Clusters Not Detected in F547M or F814W \label{tab:uphot}}
\tablehead{
   \colhead{ID} & \colhead{$\alpha$(J2000)} & \colhead{$\delta$(J2000)} & 
	\colhead{$m_{F300W}$} & \colhead{$m_{F547M}$} & 
        \colhead{E(B-V)} & \colhead{EW(H$\alpha$)} \\

   \colhead{ } & \colhead{ } & \colhead{ } & 
   \colhead{(STMAG)} & \colhead{(STMAG)} & \colhead{(mag)} & 
   \colhead{(\AA)} \\
}
\startdata
  1 & 13:36:59.61 & -29:51:55.28 &  19.65 &  20.31 &  0.00   &  \nodata    \\
  2 & 13:36:59.96 & -29:51:46.35 &  18.42 &  19.20 &  0.00   &  113\tnm{a} \\
  3 & 13:37:00.00 & -29:51:47.80 &  19.40 &  19.29 &  0.00   &  \nodata    \\
  4 & 13:37:00.09 & -29:52:01.77 &  19.08 &  19.69 &  0.00   &  224\tnm{a} \\
  5 & 13:37:00.10 & -29:51:51.00 &  18.34 &  19.87 &  0.00   &  \nodata    \\
  6 & 13:37:00.23 & -29:52:06.34 &  20.05 &  20.40 &  0.00   &  \nodata    \\
  7 & 13:37:00.34 & -29:51:58.11 &  17.52 &  18.62 &  0.50   &  \nodata    \\
  8 & 13:37:00.44 & -29:51:59.75 &  15.83 &  17.32 &  0.14   &  \nodata    \\
  9 & 13:37:00.47 & -29:51:58.82 &  17.97 &  19.38 &  0.00   &  \nodata    \\
 10 & 13:37:00.48 & -29:51:55.62 &  16.20 &  18.71 &  0.43   &  \nodata    \\
 11 & 13:37:00.48 & -29:51:59.05 &  18.29 &  19.84 &  0.00   &  \nodata    \\
 12 & 13:37:00.48 & -29:51:54.62 &  16.56 &  18.16 &  0.34   &  158\tnm{a} \\
 13 & 13:37:00.49 & -29:52:05.13 &  18.39 &  19.03 &  0.00   &  \nodata    \\
 14 & 13:37:00.50 & -29:51:55.23 &  16.61 &  18.49 &  0.17   &  757        \\
 15 & 13:37:00.52 & -29:51:56.26 &  17.15 &  18.75 &  0.33   &  \nodata    \\
 16 & 13:37:00.52 & -29:52:00.75 &  17.87 &  19.47 &  0.00   &  \nodata    \\
 17 & 13:37:00.54 & -29:52:09.46 &  20.40 &  20.39 &  0.00   &  \nodata    \\
 18 & 13:37:00.55 & -29:52:09.76 &  19.09 &  20.22 &  0.00   &  \nodata    \\
 19 & 13:37:00.55 & -29:52:00.04 &  18.50 &  20.75 &  0.00   &  \nodata    \\
 20 & 13:37:00.55 & -29:52:02.37 &  17.36 &  19.48 &  0.16   &  \nodata    \\
 21 & 13:37:00.56 & -29:51:56.49 &  17.50 &  19.07 &  0.51   &  \nodata    \\
 22 & 13:37:00.58 & -29:51:55.59 &  16.42 &  18.63 &  0.55   &  \nodata    \\
 23 & 13:37:00.59 & -29:51:59.01 &  18.89 &  19.20 &  0.00   &  \nodata    \\
 24 & 13:37:00.61 & -29:52:00.75 &  18.41 &  20.21 &  0.00   &  \nodata    \\
 25 & 13:37:00.61 & -29:51:59.36 &  18.42 &  20.04 &  0.00   &  \nodata    \\
 26 & 13:37:00.63 & -29:51:58.98 &  17.86 &  19.28 &  0.09   &  \nodata    \\
 27 & 13:37:00.63 & -29:51:59.58 &  18.33 &  19.12 &  0.00   &  \nodata    \\
 28 & 13:37:00.64 & -29:51:59.31 &  18.32 &  19.60 &  0.00   &  \nodata    \\
 29 & 13:37:00.66 & -29:52:01.70 &  18.56 &  20.01 &  0.00   &  \nodata    \\
 30 & 13:37:00.66 & -29:51:59.73 &  17.77 &  19.64 &  0.00   &  \nodata    \\
 31 & 13:37:00.68 & -29:51:58.39 &  17.07 &  18.88 &  0.36   &  \nodata    \\
 32 & 13:37:00.69 & -29:52:00.16 &  18.49 &  19.78 &  0.00   &  \nodata    \\
 33 & 13:37:00.70 & -29:52:00.59 &  18.16 &  19.43 &  0.00   &  \nodata    \\
 34 & 13:37:00.70 & -29:51:58.06 &  16.67 &  18.28 &  0.68   &  \nodata    \\
 35 & 13:37:00.71 & -29:52:00.34 &  17.97 &  19.69 &  0.00   &  \nodata    \\
 36 & 13:37:00.72 & -29:51:58.34 &  17.11 &  18.65 &  0.35   &  \nodata    \\
 37 & 13:37:00.72 & -29:52:00.07 &  18.39 &  19.72 &  0.00   &  \nodata    \\
 38 & 13:37:00.73 & -29:51:56.53 &  17.69 &  19.02 &  0.00   &  \nodata    \\
 39 & 13:37:00.76 & -29:51:57.06 &  17.91 &  18.92 &  0.00   &  \nodata    \\
 40 & 13:37:00.77 & -29:51:56.80 &  18.28 &  19.09 &  0.00   &  \nodata    \\
 41 & 13:37:00.77 & -29:51:58.31 &  17.69 &  18.79 &  0.49   &  \nodata    \\
 42 & 13:37:00.78 & -29:51:57.00 &  18.05 &  19.32 &  0.00   &  \nodata    \\
 43 & 13:37:00.78 & -29:51:57.52 &  16.46 &  18.59 &  0.28   &  \nodata    \\
 44 & 13:37:00.78 & -29:51:56.66 &  17.97 &  19.12 &  0.00   &  \nodata    \\
 45 & 13:37:00.80 & -29:51:57.96 &  16.98 &  18.75 &  0.54   &  \nodata    \\
 46 & 13:37:00.80 & -29:51:57.31 &  17.87 &  19.21 &  0.11   &  \nodata    \\
 47 & 13:37:00.80 & -29:51:56.90 &  18.02 &  19.14 &  0.06   &  \nodata    \\
 48 & 13:37:00.80 & -29:52:02.80 &  18.18 &  19.99 &  0.00   &  \nodata    \\
 49 & 13:37:00.87 & -29:52:03.06 &  18.76 &  20.19 &  0.00   &  \nodata    \\
 50 & 13:37:00.89 & -29:52:03.93 &  18.34 &  19.99 &  0.00   &  \nodata    \\
 51 & 13:37:00.90 & -29:51:57.20 &  17.85 &  18.89 &  0.17   &  \nodata    \\
 52 & 13:37:00.92 & -29:51:55.72 &  18.13 &  16.49 &  0.00   &   89        \\
 53 & 13:37:00.94 & -29:52:04.50 &  18.38 &  19.90 &  0.00   &  \nodata    \\
 54 & 13:37:00.96 & -29:52:02.15 &  18.53 &  20.46 &  0.00   &  \nodata    \\
 55 & 13:37:00.97 & -29:52:02.79 &  19.03 &  19.86 &  0.00   &  \nodata    \\
 56 & 13:37:00.97 & -29:52:02.53 &  18.32 &  19.66 &  0.00   &  \nodata    \\
 57 & 13:37:00.97 & -29:52:01.00 &  20.64 &  20.34 &  0.00   &  \nodata    \\
 58 & 13:37:01.01 & -29:52:02.98 &  17.47 &  19.47 &  0.00   &  \nodata    \\
 59 & 13:37:01.04 & -29:52:02.46 &  17.66 &  19.81 &  0.00   &  \nodata    \\
 60 & 13:37:01.05 & -29:52:03.26 &  17.52 &  18.61 &  0.00   &   29        \\
 61 & 13:37:01.09 & -29:51:57.43 &  18.08 &  19.33 &  0.12   &  \nodata    \\
 62 & 13:37:01.16 & -29:51:58.99 &  15.91 &  18.64 &  0.81   &  \nodata    \\
 63 & 13:37:01.17 & -29:51:58.50 &  16.63 &  18.12 &  0.79   &  \nodata    \\
 64 & 13:37:02.04 & -29:52:08.94 &  19.55 &  20.38 &  0.00   &  \nodata    \\
 65 & 13:36:59.55 & -29:51:59.03 &  20.00 &  \nodata &  0.00   &  \nodata    \\
 66 & 13:36:59.67 & -29:51:47.37 &  20.25 &  \nodata &  0.00   &  \nodata    \\
 67 & 13:36:59.70 & -29:51:53.32 &  19.78 &  \nodata &  0.00   &  \nodata    \\
 68 & 13:36:59.79 & -29:51:54.70 &  18.43 &  \nodata &  0.00   &  \nodata    \\
 69 & 13:36:59.87 & -29:51:58.63 &  20.17 &  \nodata &  0.00   &  \nodata    \\
 70 & 13:36:59.89 & -29:51:49.46 &  19.45 &  \nodata &  0.16   &  \nodata    \\
 71 & 13:36:59.96 & -29:51:48.29 &  19.33 &  \nodata &  0.00   &  \nodata    \\
 72 & 13:36:59.99 & -29:51:47.71 &  19.36 &  \nodata &  0.00   &  \nodata    \\
 73 & 13:37:00.05 & -29:51:50.81 &  19.52 &  \nodata &  0.00   &  \nodata    \\
 74 & 13:37:00.09 & -29:51:50.26 &  19.38 &  \nodata &  0.22   &  \nodata    \\
 75 & 13:37:00.10 & -29:51:50.76 &  18.58 &  \nodata &  0.00   &  \nodata    \\
 76 & 13:37:00.10 & -29:51:51.34 &  19.23 &  \nodata &  0.00   &  \nodata    \\
 77 & 13:37:00.12 & -29:51:50.78 &  19.26 &  \nodata &  0.00   &  \nodata    \\
 78 & 13:37:00.13 & -29:51:51.26 &  18.68 &  \nodata &  0.00   &  \nodata    \\
 79 & 13:37:00.14 & -29:51:51.49 &  17.85 &  \nodata &  0.53   &  \nodata    \\
 80 & 13:37:00.15 & -29:51:50.68 &  17.89 &  \nodata &  0.42   &  \nodata    \\
 81 & 13:37:00.18 & -29:51:51.11 &  18.14 &  \nodata &  0.76   &  834        \\
 82 & 13:37:00.19 & -29:51:44.22 &  19.90 &  \nodata &  0.00   &  \nodata    \\
 83 & 13:37:00.35 & -29:52:00.78 &  18.53 &  \nodata &  0.37   &  \nodata    \\
 84 & 13:37:00.37 & -29:51:58.52 &  17.67 &  \nodata &  0.39   &  \nodata    \\
 85 & 13:37:00.38 & -29:51:58.55 &  17.72 &  \nodata &  0.41   &  \nodata    \\
 86 & 13:37:00.38 & -29:51:59.24 &  18.01 &  \nodata &  0.11   &  \nodata    \\
 87 & 13:37:00.39 & -29:51:57.65 &  18.30 &  \nodata &  0.21   &  \nodata    \\
 88 & 13:37:00.39 & -29:51:58.19 &  17.66 &  \nodata &  0.01   &  \nodata    \\
 89 & 13:37:00.40 & -29:51:56.85 &  17.89 &  \nodata &  0.34   &  \nodata    \\
 90 & 13:37:00.40 & -29:51:58.00 &  17.74 &  \nodata &  0.11   &  \nodata    \\
 91 & 13:37:00.42 & -29:51:59.66 &  17.71 &  \nodata &  0.23   &  \nodata    \\
 92 & 13:37:00.42 & -29:51:57.55 &  18.21 &  \nodata &  0.37   &  \nodata    \\
 93 & 13:37:00.42 & -29:51:57.00 &  18.42 &  \nodata &  0.17   &  \nodata    \\
 94 & 13:37:00.43 & -29:51:57.23 &  18.45 &  \nodata &  0.26   &  \nodata    \\
 95 & 13:37:00.43 & -29:51:59.08 &  18.15 &  \nodata &  0.18   &  \nodata    \\
 96 & 13:37:00.43 & -29:51:59.53 &  17.31 &  \nodata &  0.12   &  \nodata    \\
 97 & 13:37:00.43 & -29:51:58.19 &  17.22 &  \nodata &  0.30   &  397        \\
 98 & 13:37:00.43 & -29:52:00.11 &  18.82 &  \nodata &  0.19   &  \nodata    \\
 99 & 13:37:00.44 & -29:51:56.97 &  18.98 &  \nodata &  0.00   &  \nodata    \\
100 & 13:37:00.44 & -29:51:57.51 &  19.83 &  \nodata &  0.04   &  \nodata    \\
101 & 13:37:00.44 & -29:52:00.99 &  17.90 &  \nodata &  0.31   &  \nodata    \\
102 & 13:37:00.45 & -29:51:58.70 &  18.14 &  \nodata &  0.00   &  \nodata    \\
103 & 13:37:00.45 & -29:51:58.94 &  17.93 &  \nodata &  0.11   &  \nodata    \\
104 & 13:37:00.45 & -29:51:55.60 &  17.15 &  \nodata &  0.58   &  \nodata    \\
105 & 13:37:00.45 & -29:51:58.04 &  18.45 &  \nodata &  0.01   &  \nodata    \\
106 & 13:37:00.45 & -29:51:59.40 &  17.86 &  \nodata &  0.00   &  \nodata    \\
107 & 13:37:00.46 & -29:52:01.26 &  18.92 &  \nodata &  0.28   &  \nodata    \\
108 & 13:37:00.47 & -29:51:58.65 &  18.38 &  \nodata &  0.00   &  \nodata    \\
109 & 13:37:00.47 & -29:51:59.70 &  17.19 &  \nodata &  0.00   &  \nodata    \\
110 & 13:37:00.47 & -29:51:54.37 &  16.57 &  \nodata &  0.91   &  3120       \\
111 & 13:37:00.47 & -29:51:55.02 &  17.32 &  \nodata &  0.10   &  \nodata    \\
112 & 13:37:00.47 & -29:51:59.43 &  18.45 &  \nodata &  0.00   &  \nodata    \\
113 & 13:37:00.47 & -29:52:00.18 &  16.74 &  \nodata &  0.34   &  \nodata    \\
114 & 13:37:00.48 & -29:51:56.57 &  18.05 &  \nodata &  0.42   &  \nodata    \\
115 & 13:37:00.48 & -29:51:55.53 &  17.11 &  \nodata &  0.25   &  \nodata    \\
116 & 13:37:00.48 & -29:51:54.93 &  16.77 &  \nodata &  0.22   &  \nodata    \\
117 & 13:37:00.48 & -29:52:00.59 &  17.65 &  \nodata &  0.26   &  \nodata    \\
118 & 13:37:00.48 & -29:51:54.62 &  16.56 &  \nodata &  0.34   &  \nodata    \\
119 & 13:37:00.48 & -29:51:58.23 &  19.06 &  \nodata &  0.01   &  \nodata    \\
120 & 13:37:00.49 & -29:51:59.27 &  17.80 &  \nodata &  0.00   &  210        \\
121 & 13:37:00.49 & -29:52:00.03 &  17.86 &  \nodata &  0.13   &  \nodata    \\
122 & 13:37:00.49 & -29:51:59.80 &  18.08 &  \nodata &  0.01   &  \nodata    \\
123 & 13:37:00.49 & -29:51:59.53 &  18.36 &  \nodata &  0.00   &  \nodata    \\
124 & 13:37:00.50 & -29:51:59.77 &  18.21 &  \nodata &  0.00   &  \nodata    \\
125 & 13:37:00.50 & -29:52:00.32 &  18.14 &  \nodata &  0.08   &  \nodata    \\
126 & 13:37:00.50 & -29:52:01.44 &  19.16 &  \nodata &  0.04   &  \nodata    \\
127 & 13:37:00.50 & -29:51:54.67 &  16.59 &  \nodata &  0.35   &  \nodata    \\
128 & 13:37:00.50 & -29:51:59.02 &  18.36 &  \nodata &  0.00   &  \nodata    \\
129 & 13:37:00.50 & -29:51:58.58 &  19.03 &  \nodata &  0.00   &  \nodata    \\
130 & 13:37:00.50 & -29:52:00.70 &  18.12 &  \nodata &  0.00   &  \nodata    \\
131 & 13:37:00.50 & -29:51:55.44 &  17.31 &  \nodata &  0.30   &  \nodata    \\
132 & 13:37:00.51 & -29:51:59.55 &  17.99 &  \nodata &  0.00   &  \nodata    \\
133 & 13:37:00.51 & -29:52:00.50 &  18.02 &  \nodata &  0.00   &  \nodata    \\
134 & 13:37:00.51 & -29:52:00.80 &  18.11 &  \nodata &  0.00   &  \nodata    \\
135 & 13:37:00.52 & -29:51:55.54 &  16.60 &  \nodata &  0.50   &  \nodata    \\
136 & 13:37:00.52 & -29:51:59.13 &  19.36 &  \nodata &  0.00   &  \nodata    \\
137 & 13:37:00.52 & -29:51:54.33 &  17.52 &  \nodata &  0.51   &  \nodata    \\
138 & 13:37:00.52 & -29:52:00.27 &  18.14 &  \nodata &  0.00   &  \nodata    \\
139 & 13:37:00.52 & -29:51:55.81 &  17.33 &  \nodata &  0.58   &  \nodata    \\
140 & 13:37:00.52 & -29:52:01.05 &  18.34 &  \nodata &  0.00   &  \nodata    \\
141 & 13:37:00.52 & -29:52:01.42 &  19.65 &  \nodata &  0.00   &  \nodata    \\
142 & 13:37:00.53 & -29:51:55.17 &  17.23 &  \nodata &  0.45   &  \nodata    \\
143 & 13:37:00.53 & -29:51:54.62 &  17.16 &  \nodata &  0.45   &  \nodata    \\
144 & 13:37:00.53 & -29:51:59.56 &  19.63 &  \nodata &  0.00   &  \nodata    \\
145 & 13:37:00.54 & -29:52:01.75 &  17.98 &  \nodata &  0.24   &  \nodata    \\
146 & 13:37:00.54 & -29:51:58.89 &  19.60 &  \nodata &  0.00   &  \nodata    \\
147 & 13:37:00.54 & -29:51:54.79 &  16.77 &  \nodata &  0.44   &  \nodata    \\
148 & 13:37:00.54 & -29:52:00.01 &  18.68 &  \nodata &  0.00   &  \nodata    \\
149 & 13:37:00.54 & -29:52:00.86 &  17.81 &  \nodata &  0.00   &  \nodata    \\
150 & 13:37:00.54 & -29:52:00.39 &  18.65 &  \nodata &  0.00   &  \nodata    \\
151 & 13:37:00.55 & -29:51:59.21 &  18.98 &  \nodata &  0.00   &  \nodata    \\
152 & 13:37:00.55 & -29:52:00.56 &  17.75 &  \nodata &  0.00   &  \nodata    \\
153 & 13:37:00.55 & -29:51:58.67 &  18.39 &  \nodata &  0.00   &  \nodata    \\
154 & 13:37:00.56 & -29:51:58.87 &  18.74 &  \nodata &  0.00   &  \nodata    \\
155 & 13:37:00.56 & -29:51:59.63 &  18.80 &  \nodata &  0.00   &  \nodata    \\
156 & 13:37:00.56 & -29:52:00.60 &  19.41 &  \nodata &  0.00   &  \nodata    \\
157 & 13:37:00.56 & -29:51:53.42 &  14.02 &  \nodata &  1.32   &  \nodata    \\
158 & 13:37:00.57 & -29:51:55.21 &  17.11 &  \nodata &  0.54   &  \nodata    \\
159 & 13:37:00.57 & -29:51:59.94 &  19.17 &  \nodata &  0.00   &  \nodata    \\
160 & 13:37:00.57 & -29:52:00.86 &  17.67 &  \nodata &  0.00   &   14        \\
161 & 13:37:00.57 & -29:51:59.09 &  19.54 &  \nodata &  0.00   &  \nodata    \\
162 & 13:37:00.58 & -29:51:58.70 &  18.48 &  \nodata &  0.00   &  \nodata    \\
163 & 13:37:00.58 & -29:52:00.22 &  18.56 &  \nodata &  0.00   &  \nodata    \\
164 & 13:37:00.58 & -29:51:59.35 &  19.54 &  \nodata &  0.00   &  \nodata    \\
165 & 13:37:00.58 & -29:52:00.44 &  18.91 &  \nodata &  0.00   &  \nodata    \\
166 & 13:37:00.59 & -29:52:01.50 &  17.27 &  \nodata &  0.00   &  \nodata    \\
167 & 13:37:00.59 & -29:52:01.95 &  18.24 &  \nodata &  0.12   &  \nodata    \\
168 & 13:37:00.59 & -29:52:00.69 &  18.81 &  \nodata &  0.00   &  \nodata    \\
169 & 13:37:00.59 & -29:51:59.83 &  19.05 &  \nodata &  0.00   &  \nodata    \\
170 & 13:37:00.60 & -29:52:00.24 &  18.60 &  \nodata &  0.00   &  \nodata    \\
171 & 13:37:00.60 & -29:52:01.75 &  17.54 &  \nodata &  0.00   &  \nodata    \\
172 & 13:37:00.60 & -29:51:59.47 &  19.64 &  \nodata &  0.00   &  \nodata    \\
173 & 13:37:00.60 & -29:51:55.60 &  16.81 &  \nodata &  0.61   &  \nodata    \\
174 & 13:37:00.60 & -29:52:03.64 &  20.30 &  \nodata &  0.22   &  \nodata    \\
175 & 13:37:00.61 & -29:52:01.58 &  18.41 &  \nodata &  0.18   &  \nodata    \\
176 & 13:37:00.61 & -29:51:59.81 &  19.06 &  \nodata &  0.00   &  \nodata    \\
177 & 13:37:00.61 & -29:52:04.54 &  21.02 &  \nodata &  0.00   &  \nodata    \\
178 & 13:37:00.62 & -29:52:01.42 &  17.50 &  \nodata &  0.00   &  \nodata    \\
179 & 13:37:00.62 & -29:51:55.56 &  17.08 &  \nodata &  0.61   &  \nodata    \\
180 & 13:37:00.62 & -29:51:58.80 &  17.65 &  \nodata &  0.15   &  \nodata    \\
181 & 13:37:00.62 & -29:52:00.30 &  18.84 &  \nodata &  0.00   &  \nodata    \\
182 & 13:37:00.62 & -29:52:01.66 &  17.34 &  \nodata &  0.12   &  \nodata    \\
183 & 13:37:00.63 & -29:52:00.60 &  19.66 &  \nodata &  0.00   &  \nodata    \\
184 & 13:37:00.63 & -29:52:00.92 &  18.76 &  \nodata &  0.00   &  \nodata    \\
185 & 13:37:00.63 & -29:52:01.21 &  17.94 &  \nodata &  0.00   &  \nodata    \\
186 & 13:37:00.63 & -29:52:00.03 &  18.74 &  \nodata &  0.00   &  \nodata    \\
187 & 13:37:00.63 & -29:52:03.73 &  19.38 &  \nodata &  0.16   &  \nodata    \\
188 & 13:37:00.63 & -29:51:59.69 &  18.44 &  \nodata &  0.00   &  \nodata    \\
189 & 13:37:00.64 & -29:51:59.72 &  18.47 &  \nodata &  0.00   &  \nodata    \\
190 & 13:37:00.64 & -29:52:00.24 &  18.76 &  \nodata &  0.00   &  \nodata    \\
191 & 13:37:00.65 & -29:51:58.11 &  16.85 &  \nodata &  0.60   &  \nodata    \\
192 & 13:37:00.65 & -29:52:01.30 &  19.07 &  \nodata &  0.00   &  \nodata    \\
193 & 13:37:00.66 & -29:51:59.45 &  18.38 &  \nodata &  0.00   &  \nodata    \\
194 & 13:37:00.66 & -29:52:00.11 &  18.54 &  \nodata &  0.00   &  \nodata    \\
195 & 13:37:00.66 & -29:51:56.65 &  16.74 &  \nodata &  0.94   &  559        \\
196 & 13:37:00.66 & -29:52:03.15 &  19.21 &  \nodata &  0.00   &  \nodata    \\
197 & 13:37:00.67 & -29:52:01.84 &  18.86 &  \nodata &  0.00   &  \nodata    \\
198 & 13:37:00.67 & -29:52:02.87 &  18.65 &  \nodata &  0.00   &  \nodata    \\
199 & 13:37:00.68 & -29:51:58.28 &  18.18 &  \nodata &  0.28   &  \nodata    \\
200 & 13:37:00.68 & -29:52:00.51 &  18.98 &  \nodata &  0.00   &  \nodata    \\
201 & 13:37:00.68 & -29:52:00.81 &  19.21 &  \nodata &  0.01   &  \nodata    \\
202 & 13:37:00.68 & -29:52:00.16 &  18.32 &  \nodata &  0.00   &  \nodata    \\
203 & 13:37:00.69 & -29:51:59.73 &  18.74 &  \nodata &  0.00   &  \nodata    \\
204 & 13:37:00.69 & -29:52:02.36 &  19.41 &  \nodata &  0.05   &  \nodata    \\
205 & 13:37:00.69 & -29:51:57.20 &  18.08 &  \nodata &  0.62   &  \nodata    \\
206 & 13:37:00.69 & -29:52:01.79 &  19.47 &  \nodata &  0.00   &  \nodata    \\
207 & 13:37:00.70 & -29:52:01.38 &  18.89 &  \nodata &  0.00   &  \nodata    \\
208 & 13:37:00.70 & -29:52:04.93 &  19.35 &  \nodata &  0.00   &  \nodata    \\
209 & 13:37:00.70 & -29:52:00.96 &  19.05 &  \nodata &  0.00   &  \nodata    \\
210 & 13:37:00.71 & -29:51:56.49 &  16.67 &  \nodata &  0.61   &  \nodata    \\
211 & 13:37:00.71 & -29:52:02.62 &  19.79 &  \nodata &  0.00   &  \nodata    \\
212 & 13:37:00.71 & -29:52:01.98 &  19.36 &  \nodata &  0.00   &  \nodata    \\
213 & 13:37:00.72 & -29:52:01.71 &  19.12 &  \nodata &  0.00   &  \nodata    \\
214 & 13:37:00.72 & -29:51:56.75 &  17.82 &  \nodata &  0.26   &  \nodata    \\
215 & 13:37:00.72 & -29:52:03.02 &  21.01 &  \nodata &  0.08   &  \nodata    \\
216 & 13:37:00.72 & -29:51:57.25 &  17.45 &  \nodata &  0.66   &  \nodata    \\
217 & 13:37:00.72 & -29:52:00.65 &  18.80 &  \nodata &  0.00   &  \nodata    \\
218 & 13:37:00.72 & -29:52:02.25 &  19.21 &  \nodata &  0.00   &  \nodata    \\
219 & 13:37:00.73 & -29:52:01.87 &  18.59 &  \nodata &  0.00   &  \nodata    \\
220 & 13:37:00.73 & -29:52:01.60 &  19.68 &  \nodata &  0.00   &  \nodata    \\
221 & 13:37:00.73 & -29:52:02.61 &  18.70 &  \nodata &  0.00   &  \nodata    \\
222 & 13:37:00.74 & -29:51:56.57 &  18.43 &  \nodata &  0.00   &  \nodata    \\
223 & 13:37:00.75 & -29:51:56.40 &  18.13 &  \nodata &  0.04   &  \nodata    \\
224 & 13:37:00.75 & -29:51:56.96 &  17.34 &  \nodata &  0.37   &  \nodata    \\
225 & 13:37:00.75 & -29:52:02.41 &  20.08 &  \nodata &  0.00   &  \nodata    \\
226 & 13:37:00.75 & -29:51:56.11 &  16.70 &  \nodata &  0.54   &  \nodata    \\
227 & 13:37:00.75 & -29:51:58.30 &  17.33 &  \nodata &  0.42   &  \nodata    \\
228 & 13:37:00.76 & -29:52:02.06 &  19.62 &  \nodata &  0.00   &  \nodata    \\
229 & 13:37:00.76 & -29:51:57.50 &  17.11 &  \nodata &  0.43   &  \nodata    \\
230 & 13:37:00.76 & -29:52:01.70 &  19.52 &  \nodata &  0.00   &  \nodata    \\
231 & 13:37:00.77 & -29:51:56.49 &  18.25 &  \nodata &  0.00   &  \nodata    \\
232 & 13:37:00.77 & -29:51:58.64 &  17.78 &  \nodata &  0.36   &  \nodata    \\
233 & 13:37:00.77 & -29:52:04.90 &  19.15 &  \nodata &  0.00   &  \nodata    \\
234 & 13:37:00.77 & -29:52:00.92 &  19.24 &  \nodata &  0.06   &  \nodata    \\
235 & 13:37:00.78 & -29:51:56.21 &  18.20 &  \nodata &  0.00   &  \nodata    \\
236 & 13:37:00.78 & -29:51:57.83 &  18.14 &  \nodata &  0.45   &  \nodata    \\
237 & 13:37:00.78 & -29:52:01.20 &  20.06 &  \nodata &  0.00   &  \nodata    \\
238 & 13:37:00.79 & -29:52:02.27 &  18.78 &  \nodata &  0.00   &  \nodata    \\
239 & 13:37:00.80 & -29:51:58.38 &  18.67 &  \nodata &  0.47   &  \nodata    \\
240 & 13:37:00.80 & -29:52:02.06 &  18.95 &  \nodata &  0.00   &  \nodata    \\
241 & 13:37:00.81 & -29:51:47.10 &  20.07 &  \nodata &  0.00   &  \nodata    \\
242 & 13:37:00.81 & -29:52:03.06 &  18.95 &  \nodata &  0.00   &  \nodata    \\
243 & 13:37:00.82 & -29:52:05.67 &  20.60 &  \nodata &  0.00   &  \nodata    \\
244 & 13:37:00.82 & -29:52:02.58 &  19.10 &  \nodata &  0.00   &  \nodata    \\
245 & 13:37:00.82 & -29:51:57.51 &  18.08 &  \nodata &  0.35   &  \nodata    \\
246 & 13:37:00.82 & -29:52:02.33 &  19.08 &  \nodata &  0.00   &  \nodata    \\
247 & 13:37:00.83 & -29:52:01.39 &  19.66 &  \nodata &  0.00   &  \nodata    \\
248 & 13:37:00.83 & -29:52:03.03 &  19.22 &  \nodata &  0.00   &  \nodata    \\
249 & 13:37:00.84 & -29:52:01.63 &  18.86 &  \nodata &  0.00   &  \nodata    \\
250 & 13:37:00.84 & -29:52:01.97 &  19.36 &  \nodata &  0.00   &  \nodata    \\
251 & 13:37:00.84 & -29:52:02.48 &  19.30 &  \nodata &  0.00   &  \nodata    \\
252 & 13:37:00.85 & -29:52:01.05 &  19.45 &  \nodata &  0.00   &  \nodata    \\
253 & 13:37:00.85 & -29:51:57.69 &  18.21 &  \nodata &  0.22   &  \nodata    \\
254 & 13:37:00.85 & -29:52:06.04 &  19.37 &  \nodata &  0.26   &  \nodata    \\
255 & 13:37:00.85 & -29:52:03.14 &  18.95 &  \nodata &  0.00   &  \nodata    \\
256 & 13:37:00.86 & -29:51:57.88 &  18.08 &  \nodata &  0.32   &  \nodata    \\
257 & 13:37:00.86 & -29:52:02.05 &  19.09 &  \nodata &  0.00   &  \nodata    \\
258 & 13:37:00.87 & -29:52:01.48 &  19.33 &  \nodata &  0.00   &  \nodata    \\
259 & 13:37:00.87 & -29:52:03.30 &  18.83 &  \nodata &  0.00   &  \nodata    \\
260 & 13:37:00.87 & -29:52:03.81 &  19.21 &  \nodata &  0.00   &  \nodata    \\
261 & 13:37:00.88 & -29:51:57.48 &  18.57 &  \nodata &  0.14   &  \nodata    \\
262 & 13:37:00.88 & -29:52:02.68 &  18.19 &  \nodata &  0.00   &  \nodata    \\
263 & 13:37:00.88 & -29:52:04.79 &  18.54 &  \nodata &  0.00   &  \nodata    \\
264 & 13:37:00.89 & -29:52:03.58 &  19.37 &  \nodata &  0.00   &  \nodata    \\
265 & 13:37:00.89 & -29:52:01.77 &  20.32 &  \nodata &  0.00   &  \nodata    \\
266 & 13:37:00.89 & -29:51:59.65 &  19.47 &  \nodata &  0.00   &  \nodata    \\
267 & 13:37:00.90 & -29:52:01.51 &  19.32 &  \nodata &  0.00   &  \nodata    \\
268 & 13:37:00.90 & -29:52:04.29 &  18.70 &  \nodata &  0.00   &  \nodata    \\
269 & 13:37:00.90 & -29:52:02.52 &  18.71 &  \nodata &  0.00   &  \nodata    \\
270 & 13:37:00.90 & -29:52:03.19 &  19.22 &  \nodata &  0.00   &  \nodata    \\
271 & 13:37:00.91 & -29:52:01.25 &  19.71 &  \nodata &  0.00   &  \nodata    \\
272 & 13:37:00.91 & -29:51:59.58 &  18.29 &  \nodata &  0.20   &  \nodata    \\
273 & 13:37:00.91 & -29:52:01.90 &  19.58 &  \nodata &  0.00   &  \nodata    \\
274 & 13:37:00.92 & -29:52:03.90 &  18.57 &  \nodata &  0.00   &  \nodata    \\
275 & 13:37:00.92 & -29:52:03.21 &  18.75 &  \nodata &  0.00   &  \nodata    \\
276 & 13:37:00.92 & -29:52:04.24 &  18.67 &  \nodata &  0.00   &  \nodata    \\
277 & 13:37:00.92 & -29:52:04.73 &  18.15 &  \nodata &  0.21   &  \nodata    \\
278 & 13:37:00.93 & -29:52:03.53 &  19.58 &  \nodata &  0.00   &  \nodata    \\
279 & 13:37:00.93 & -29:52:00.12 &  19.01 &  \nodata &  0.00   &  \nodata    \\
280 & 13:37:00.93 & -29:52:02.88 &  19.29 &  \nodata &  0.00   &  \nodata    \\
281 & 13:37:00.94 & -29:52:01.14 &  19.50 &  \nodata &  0.00   &  \nodata    \\
282 & 13:37:00.94 & -29:51:59.66 &  19.61 &  \nodata &  0.01   &  \nodata    \\
283 & 13:37:00.94 & -29:52:00.83 &  19.17 &  \nodata &  0.00   &  \nodata    \\
284 & 13:37:00.95 & -29:52:03.18 &  19.17 &  \nodata &  0.00   &  \nodata    \\
285 & 13:37:00.96 & -29:52:00.18 &  19.11 &  \nodata &  0.19   &  \nodata    \\
286 & 13:37:00.96 & -29:52:01.21 &  18.62 &  \nodata &  0.09   &  \nodata    \\
287 & 13:37:00.96 & -29:51:59.83 &  18.79 &  \nodata &  0.19   &  \nodata    \\
288 & 13:37:00.97 & -29:52:02.20 &  19.18 &  \nodata &  0.00   &  \nodata    \\
289 & 13:37:00.98 & -29:52:02.82 &  18.94 &  \nodata &  0.00   &  \nodata    \\
290 & 13:37:00.98 & -29:52:04.17 &  19.24 &  \nodata &  0.00   &  \nodata    \\
291 & 13:37:00.98 & -29:52:02.41 &  18.27 &  \nodata &  0.00   &  \nodata    \\
292 & 13:37:00.98 & -29:52:03.05 &  17.99 &  \nodata &  0.00   &  \nodata    \\
293 & 13:37:00.99 & -29:52:03.42 &  18.62 &  \nodata &  0.00   &  \nodata    \\
294 & 13:37:00.99 & -29:52:04.02 &  19.44 &  \nodata &  0.00   &  \nodata    \\
295 & 13:37:00.99 & -29:52:01.01 &  19.73 &  \nodata &  0.00   &  \nodata    \\
296 & 13:37:01.00 & -29:52:03.21 &  17.77 &  \nodata &  0.00   &  \nodata    \\
297 & 13:37:01.01 & -29:52:02.13 &  18.44 &  \nodata &  0.00   &  \nodata    \\
298 & 13:37:01.01 & -29:52:02.54 &  17.92 &  \nodata &  0.00   &  \nodata    \\
299 & 13:37:01.01 & -29:52:03.41 &  17.90 &  \nodata &  0.00   &  \nodata    \\
300 & 13:37:01.02 & -29:52:02.40 &  18.13 &  \nodata &  0.00   &  \nodata    \\
301 & 13:37:01.02 & -29:52:03.15 &  17.56 &  \nodata &  0.00   &  \nodata    \\
302 & 13:37:01.02 & -29:52:02.64 &  18.08 &  \nodata &  0.00   &  \nodata    \\
303 & 13:37:01.03 & -29:52:02.93 &  17.66 &  \nodata &  0.00   &  \nodata    \\
304 & 13:37:01.03 & -29:52:00.05 &  19.84 &  \nodata &  0.00   &  \nodata    \\
305 & 13:37:01.03 & -29:52:03.40 &  17.80 &  \nodata &  0.00   &  \nodata    \\
306 & 13:37:01.04 & -29:52:02.14 &  18.38 &  \nodata &  0.00   &  \nodata    \\
307 & 13:37:01.04 & -29:52:03.13 &  17.43 &  \nodata &  0.00   &  \nodata    \\
308 & 13:37:01.05 & -29:52:02.74 &  17.77 &  \nodata &  0.00   &  \nodata    \\
309 & 13:37:01.05 & -29:52:02.94 &  17.75 &  \nodata &  0.00   &  \nodata    \\
310 & 13:37:01.06 & -29:52:02.55 &  18.15 &  \nodata &  0.00   &  \nodata    \\
311 & 13:37:01.06 & -29:52:03.22 &  18.36 &  \nodata &  0.00   &  \nodata    \\
312 & 13:37:01.07 & -29:52:02.29 &  18.28 &  \nodata &  0.00   &  \nodata    \\
313 & 13:37:01.07 & -29:52:01.76 &  18.30 &  \nodata &  0.02   &  \nodata    \\
314 & 13:37:01.07 & -29:52:02.75 &  17.94 &  \nodata &  0.00   &  \nodata    \\
315 & 13:37:01.08 & -29:52:02.12 &  18.30 &  \nodata &  0.00   &  \nodata    \\
316 & 13:37:01.08 & -29:51:40.85 &  20.34 &  \nodata &  0.00   &  \nodata    \\
317 & 13:37:01.08 & -29:52:01.34 &  17.39 &  \nodata &  0.24   &  \nodata    \\
318 & 13:37:01.08 & -29:52:03.00 &  18.60 &  \nodata &  0.00   &  \nodata    \\
319 & 13:37:01.11 & -29:51:59.71 &  19.13 &  \nodata &  0.19   &  \nodata    \\
320 & 13:37:01.13 & -29:51:53.46 &  20.31 &  \nodata &  0.00   &  \nodata    \\
321 & 13:37:01.13 & -29:51:59.30 &  19.28 &  \nodata &  0.06   &  \nodata    \\
322 & 13:37:01.18 & -29:51:58.62 &  17.19 &  \nodata &  0.42   &  \nodata    \\
323 & 13:37:01.21 & -29:51:55.72 &  19.36 &  \nodata &  0.00   &  \nodata    \\
324 & 13:37:01.24 & -29:51:57.44 &  18.20 &  \nodata &  0.35   &  \nodata    \\
325 & 13:37:01.25 & -29:51:56.88 &  18.73 &  \nodata &  0.26   &  \nodata    \\
326 & 13:37:01.38 & -29:52:05.44 &  20.65 &  \nodata &  0.00   &  \nodata    \\
327 & 13:37:01.38 & -29:52:10.33 &  20.50 &  \nodata &  0.00   &  \nodata    \\
328 & 13:37:01.59 & -29:51:57.47 &  19.81 &  \nodata &  0.00   &  \nodata    \\
329 & 13:37:01.70 & -29:51:40.49 &  20.25 &  \nodata &  0.00   &  \nodata    \\
330 & 13:37:02.00 & -29:51:57.36 &  20.16 &  \nodata &  0.00   &  \nodata    \\
\enddata
\tablenotetext{a}{EW(H$\alpha$) value estimated from flux measured 
through an 11-pixel aperture in the peak-subtracted F656N image.}
\end{deluxetable}

\clearpage

%%Begin Figures

\end{document}